\documentclass{iopjournal}

\usepackage{amsmath}
\usepackage{comment}
\usepackage{geometry}
\usepackage{array}
\usepackage{multirow}
\usepackage{subcaption}
\usepackage{bm}
\usepackage{xcolor}
\usepackage{hyperref}
\usepackage{ragged2e}
\usepackage[normalem]{ulem}

\hypersetup{
    colorlinks=true,
    linkcolor=black,
    citecolor=black,
    urlcolor=black
}
\begin{document}

\title{Elastic tensor-derived properties of composition-dependent disordered refractory binary alloys using DFPT}
\author{Surya T.~Bijjala$^1$\orcid{0009-0005-4805-0829}, Susan R. Atlas$^{2,*}$\orcid{0000-0003-1542-2700} and Pankaj Kumar$^{3,*}$\orcid{0000-0003-2209-3538}}

\affil{$^1$Department of Mechanical Engineering, University of New Mexico, Albuquerque, NM USA}

\affil{$^2$Department of Chemistry and Chemical Biology, Department of Physics and Astronomy, and Center for Quantum Information and Control, University of New Mexico, Albuquerque, NM USA}

\affil{$^3$Department of Mechanical Engineering, University of New Mexico, Albuquerque, NM USA}

\vspace*{.04in}

\affil{$^*$Authors to whom any correspondence should be addressed.}
\email{pankaj@unm.edu; susier@unm.edu}

\keywords{refractory alloy, density functional theory, compositionally-complex alloys, elastic properties, \\ 
\vspace*{-0.02in} DFPT, property anisotropy}

\begin{abstract}
\justifying
The elastic tensor provides valuable insight into the mechanical behavior of a material with lattice strain, such as disordered binary alloys. Traditional stress-strain methods have made it possible to compute elastic constants for high-throughput databases of ordered structures, as well as individually tailored alloy compositions. However, this approach depends on predetermined or iteratively-chosen strain tensors. This poses a significant challenge for systematic, composition-dependent studies of disordered materials 
requiring structural models with randomized atomic positions and possessing low symmetry. Density functional perturbation theory (DFPT) based on density functional theory provides a compelling alternative to stress-strain methods: it allows for an unbiased determination of the elastic tensor,  as well as access to local field data derived from the underlying general response function framework.  Despite its intrinsic flexibility and efficiency, DFPT has seen limited application to the study of disordered systems.  At the same time, there is a growing need for expanded quantum mechanical data to improve predictive modeling of complex disordered material properties. Here we present results for the rigid-ion and relaxed-ion elastic tensors computed using DFPT, for a comprehensive set of disordered structural refractory body-centered cubic (BCC) binary alloys of molybdenum (Mo), niobium (Nb), tantalum (Ta), and tungsten (W). For the first time, we have mapped the quantum-driven heterogeneity in elastic constants and derived mechanical properties---bulk modulus ($B$), shear modulus ($G$), Young’s modulus ($E$), and Poisson’s ratio ($\nu$)---and associated nuclear-relaxation fields at each disordered structure lattice site. The derived properties, together with Pugh’s ratio ($B/G$), Cauchy pressure and elastic anisotropy, are reported as a function of composition for all refractory binaries. The DFPT-computed elastic constant results for the refractory binary alloys at systematically-varied Mo, Nb, Ta, and W compositions are in excellent agreement with available experimental data. 
\end{abstract}

\justifying
\section{Introduction}
\label{sec:Introduction}
The elastic tensor \cite{dieter1976mechanical} and its elastic constant components give a complete description, in terms of deformation, of material response under external stress within the elastic limit. The elastic tensor can therefore serve as a valuable quantitative probe of elastic strain effects associated with changes in chemical composition, and their impact on mechanical \cite{liang2024first,marker2018effects,muzyk2011phase, huhn2014first,hill1952elastic}, electronic \cite{bouarissa2001mechanical, de2007ab}, thermodynamic \cite{laberge1995elastic, zhang2010effects}, and structural \cite{born1940stability, born1954dynamical, persson1999dynamical} properties in these materials. 

Refractory alloys show great potential to meet the demands for high-temperature materials due to their high melting points and ability to maintain mechanical properties at temperatures higher than current Ni-based superalloys \cite{senkov2018development}. The refractory elements Nb, Ta, Mo, and W of groups V and VI comprise the BCC structural refractory metals family, characterized by high mechanical strength and melting temperatures \cite{nieh1993recent}. Their structural refractory binary alloys MoNb, MoTa, MoW, NbTa, WNb and WTa form solid solutions of BCC crystal structure across their entire compositional space \cite{bijjala2024understanding}, and form an attractive starting point for designing more complex disordered alloys. However, elastic constant measurements are typically reported only for a limited number of alloy classes and compositions, and experiments often exhibit substantial disagreement even between values measured using the same technique and on the same material \cite{davies1977transducer, beg1976study}. Multiple factors can influence differences in reported experimental values, such as impurities and measurement temperature \cite{shang2007first, hu2021screening, borges2024local, giles2022machine}. Previous efforts have focused on the computational determination of elastic constants using density functional theory (DFT) for selected refractory binary alloys, using stress-strain methods \cite{jiang2018first, mei2023predicting, ge2017elastic, bhandari2022predicting, wang2020first}. While these investigations have provided valuable insights, they were limited to a few binary systems or computed for a narrow compositional range. The computational methodologies employed in these studies were primarily validated using experimental data for pure elements. Since the electronic interactions in alloy systems are inherently more complex than in their elemental counterparts, the direct application of methods validated for pure elements to alloy systems may introduce significant uncertainties, potentially compromising the accuracy and practical relevance of the predicted elastic properties for these alloys. These challenges are compounded when modeling disordered materials \cite{horton2025accelerated}. Here, we establish and validate a first-principles modeling framework based on DFT \cite{hohenberg1964, kohn1965} for performing computationally-efficient studies of elastic tensor and derived properties of disordered alloys, validated for systematically-varied Mo, Nb, Ta, and W binary alloy compositions. DFT, as implemented in state-of-the-art plane wave pseudopotential periodic electronic structure codes such as Abinit \cite{gonze2020abinit,romero2020abinit}, provides direct access to the computation of energies and stresses for obtaining the elastic tensors of the materials, with only atomic structure and nuclear charges as inputs.

The principal quantity to be considered is the elastic tensor and its elastic constant components. 
The stress-strain relation is given by:
\begin{equation}
\label{eq:stress-strain}
\begin{bmatrix}
\sigma_{11} \\
\sigma_{22} \\
\sigma_{33} \\
\sigma_{23} \\
\sigma_{13} \\
\sigma_{12}
\end{bmatrix}
=
\begin{bmatrix}
C_{11} & C_{12} & C_{13} & C_{14} & C_{15} & C_{16} \\
C_{12} & C_{22} & C_{23} & C_{24} & C_{25} & C_{26} \\
C_{13} & C_{23} & C_{33} & C_{34} & C_{35} & C_{36} \\
C_{14} & C_{24} & C_{34} & C_{44} & C_{45} & C_{46} \\
C_{15} & C_{25} & C_{35} & C_{45} & C_{55} & C_{56} \\
C_{16} & C_{26} & C_{36} & C_{46} & C_{56} & C_{66}
\end{bmatrix}
\begin{bmatrix}
s_{11} \\
s_{22} \\
s_{33} \\
2s_{23} \\
2s_{13} \\
2s_{12}
\end{bmatrix}
\end{equation}
where $\boldsymbol{\sigma}$ is the $6 \times 1$ stress tensor, the $C_{ij}$ are the elastic constants, and $\boldsymbol{s}$ is the $6\times 1$ strain tensor in Voigt notation \cite{simmons1971single}.

In the traditional stress-strain approach, the elastic constants $C_{ij}$ are obtained by computing either the total energy or the stress of a crystal at a series of preselected strains \cite{shang2007first,dal2016elastic,luo2012ab,shao2012temperature, steinle1999first}. The selection of strain sets can significantly influence the computed elastic constants, with variations exceeding 15\% \cite{de2015charting} in some cases. To mitigate this problem, high-throughput databases of ordered materials, such as the Materials Project \cite{Jain2013}, validate computed elastic constants by comparing results obtained from multiple strain sets \cite{de2015charting}. In disordered alloys, such as the refractory binary alloys that are the focus of the present work, this approach is not practical due to the lack of crystal symmetry caused by intrinsic lattice strain.

An alternative and more general approach that avoids the need for tailored strains is based on analytical differentiation of the energy using linear response theory \cite{baroni1987elastic, hamann2005metric,hamann2005generalized, hamann2005systematic}. In this method, the elastic tensor is obtained by expanding the total energy with respect to infinitesimal perturbation of the system using density functional perturbation theory (DFPT) \cite{baroni2001phonons}. 

DFPT has a number of important advantages in the present context: (i) a preselected list of strains as a function of material and composition is not required, which avoids the potential generation of unphysical structures for unexplored alloys; (ii) it is generally applicable to all types of complex alloy systems, including those with large unit cells or reduced crystallographic symmetry, making it particularly well-suited for studying disordered or low-symmetry alloys; (iii) in contrast to the stress-strain method, determining the DFPT elastic tensor requires only a single calculation rather than a series of strain applications; (iv) because DFPT is couched in terms of a response function framework, it provides access to local field information that cannot be obtained from the manual application of a stress-strain approach. As shown below, such maps provide quantitative measures of the microscopic elastic response of the binary alloys, for potential input to data-driven predictive models of materials properties.
 
DFPT for the elastic tensor was initially implemented for norm-conserving pseudopotentials (NCPPs) \cite{Gonze1997a} and later extended to the projector augmented-wave (PAW) method by Martin \textit{et al.}~\cite{martin2019projector}. DFPT has been successfully demonstrated using both NCPPs \cite{soumelidou2017strain, stenuit2007first, bouamama2009ab} and PAW pseudopotentials \cite{daoust2022longitudinal, daoust2021impact, ma2022investigation, zha2023alloying, peng2019tunable, jia2021origin}. Previous applications of the DFPT methodology have primarily focused on the study of piezoelectric materials \cite{daoust2022longitudinal, daoust2021impact, ma2022investigation, zha2023alloying, peng2019tunable, jia2021origin, ko2022first, zhang2025deep, zha2023enhanced, wang2022band}.
Despite its intrinsic advantages, DFPT for elastic properties has been applied to only a limited number of structural alloys, {\it e.g.}~the ternary alloys $({\rm Al}_x{\rm Mg}_{1-x}){\rm Sc}$ \cite{saha2014elastic} and alloys of aluminum nitride and rare-earth nitride \cite{daoust2022longitudinal}. The present work represents the first demonstration of DFPT for computing the elastic properties of the structural refractory disordered binary alloys of Mo, Nb, Ta and W. 

Within the DFPT framework, the elastic tensor can be evaluated as the \textit{rigid-ion} elastic tensor, $C_{ij}^{rigid}$, or as the \textit{relaxed-ion} elastic tensor $C_{ij}^{relaxed}$ \cite{baroni1987elastic, hamann2005metric,hamann2005generalized, hamann2005systematic}. $C_{ij}^{rigid}$ is computed with the relaxed initial structure, and strain as the perturbation; $C_{ij}^{relaxed}$ is a correction to the rigid-ion elastic tensor, with both strain and atomic displacements (nuclear relaxations) as the perturbations. Both $C_{ij}^{rigid}$ and $C_{ij}^{relaxed}$ provide essential information about the elastic response of a material. The relaxed-ion elastic tensor is the quantity that will correspond most closely to experiment. 

Comparison of the rigid-ion and relaxed-ion results allows an assessment of the accuracy of the rigid-ion approximation, as well as enabling the mapping of the \textit{ nuclear relaxation field} of a disordered alloy. We define the nuclear relaxation elastic tensor $C_{ij}^{\text{nuclear}}$ as the difference between the corresponding rigid-ion and relaxed-ion quantities \cite{hamann2005systematic}:
\begin{equation}
\label{eq:nuclear-relaxations}
C_{ij}^{\text{relaxed}} = C_{ij}^{\text{rigid}} - C_{ij}^{\text{nuclear}}
\end{equation}
where
\begin{equation}
\label{eq:C_rigid}
 C_{ij}^{\text{rigid}} = \frac{d\sigma_i}{ds_j} = \frac{\partial^2 E}{\partial s_i \partial s_j}.
\end{equation}
$E$ is the energy per undeformed unit cell volume, and the $s_i$ are lattice deformations.

Non-zero nuclear-relaxation contributions in binary alloys are due to lattice distortions induced by the alloying elements, leading to forces between atoms in a finite strain state. Nuclear-relaxation contributions have previously been used to understand the effect of stress on piezoelectric coefficients and elastic constants by decomposing these properties into a purely electronic and nuclear-relaxation contributions resulting from the atomic rearrangements in the strained state \cite{daoust2022longitudinal, daoust2021impact,ma2022investigation, ko2022first, zhang2025deep, zha2023alloying, zha2023enhanced, wang2022band}. In the study by Zhang \textit{et al.}~\cite{zhang2006first} and in recent works by Zha \textit{et al.}~\cite{zha2023alloying, zha2023enhanced}, the nuclear-relaxation term of the piezoelectric coefficient was further decomposed into lattice point-specific relaxation contributions in order to elucidate the microscopic origin of the piezoelectric effect. In a recent study by Zhang \textit{et al.}~\cite{zhang2025deep} specific atomic contribution to piezoelectric response in $\rm{(K, Na) NbO_3}$ was studied. Despite the methodological similarities between the computation of relaxed-ion elastic constants and relaxed-ion piezoelectric properties, the atom specific contributions from nuclear relaxation to the elastic constants have not been previously reported. In this study, for the first time, microscopic insights into the influence of alloying on the mechanical behavior of refractory binary alloys are provided through the decomposition of the nuclear relaxation contributions. This approach enables the investigation of the spatial variation of elastic constants and derived elastic moduli within the crystal structure of disordered refractory binary alloys. This allows a deeper understanding of how local atomic environments affect the overall mechanical response of these complex systems.

In the present work, we utilize DFT+DFPT to compute the single-crystal, relaxed-structure elastic tensors of the six BCC refractory binary alloys of (Mo, Nb, Ta, W)---MoNb, MoTa, MoW, NbTa, MoTa, and MoW---as a function of composition. The disordered alloys are modeled as supercells constructed using the special quasi-random structures (SQS) method   \cite{zunger1990special}. The relaxed disordered structures are then used to compute the rigid-ion and relaxed-ion elastic tensors based on DFPT. We assess the effectiveness of DFPT in providing accurate elastic tensors through detailed comparison of elastic constants with available experimental data. The composition-dependent calculations serve as an important test of the ability of the SQS method to accurately model disorder in these alloys. Mechanical properties $B$, $G$, and $E$, the elastic anisotropy and $B/G$ and $C'$ derived from the computed elastic moduli are reported as a function of composition for all refractory binaries. Computed relaxation fields for the Voigt-averaged elastic constants, shear modulus, bulk modulus, Young's modulus, and Poisson's ratio are presented for all binaries and discussed in terms of constituent element chemistry. The in-depth understanding of trends in elastic properties and anisotropy in the disordered refractory binary alloys gained in this work through the decomposition of elastic constants into electronic and nuclear contributions is expected to contribute to new microscopic alloy computational strategies for designing refractory alloys for high-temperature applications.

\section{Computational Methods}
\label{sec:CompMeth}
We used the open-source, periodic plane wave pseudopotential electronic structure code Abinit \cite{gonze2020abinit, romero2020abinit} for all DFT calculations, employing the Perdew, Burke, and Ernzerhof (PBE) generalized gradient approximation (GGA) exchange-correlation functional \cite{perdew1996generalized}, with corresponding PBE norm-conserving pseudopotentials \cite{van2018pseudodojo}. We note that while meta-GGA methods such as r2SCAN \cite{furness2020accurate} and parameterized DFT+$U$ \cite{anisimov1991band} approaches have seen growing popularity for computing the properties of ordered systems with 1--3 elements \cite{kingsbury2022performance}, the corresponding pseudopotentials are not currently available, and the accuracy of these functionals for complex disordered systems with 5+ elements has yet to be assessed.  Since the present work is intended as a starting point for scaling up to larger refractory alloy systems with increased chemical complexity, the PBE functional is utilized throughout this work, to ensure a uniform level of theory across all alloys.

The plane waves for each system were expanded up to a kinetic energy of 30 Ha, with each structure converged until the total residual differences in the potential were $< 10^{-12}$ Ha, corresponding to convergence to $10^{-6}$ bohr for lattice constants, and atomic forces converged to $<|\pm 0.01|$ mHa/bohr.  Following careful convergence studies, we adopted a $\Gamma$-centered Monkhorst–Pack \cite{monkhorst1976special} special $k$-point grid with a density of 50 $k$-points per \text{\AA} for all systems.  

Initial benchmark studies were performed on bulk Mo, Nb, Ta, and W in order to determine appropriate computational parameters for the binary alloy calculations. We computed the lattice constants, bulk moduli, cohesive energies, elastic constants, and (100), (110), (111) surface energies for elemental Mo, W, Nb, and Ta, and compared our results with available experimental data, using three different pseudopotentials: the FHI norm-conserving pseudopotentials ($d$ valence electrons only) tabulated at the Abinit website \cite{Abinit-website}, pseudo-dojo \cite{van2018pseudodojo} NCPPs ($s$, $p$ and $d$ valence electrons) available at \cite{pseudodojo-website}, and projector augmented-wave (PAW) pseudopotentials \cite{blochl1994projector} ($s$, $p$ and $d$ valence electrons), tabulated at the Abinit website \cite{Abinit-website}.  Computed properties using the FHI NCPPs showed an average deviation of $\approx 1\%$ compared to available experiments, across all computed properties for Mo, Nb and Ta, but an average deviation of $\approx 9\%$ for W. For the pseudo-dojo NCPPs and PAW-based pseudopotentials, computed properties agreed well with experiment, with an average deviation of $\approx 1.5\%$ for all elements and across all properties. Detailed results are reported in \cite{bijjala2025composition}. 
Given the comparable performance of the pseudo-dojo NCPP and PAW pseudopotentials,  the pseudo-dojo NCPPs were used throughout this work.  Working within the pseudo-dojo framework \cite{van2018pseudodojo} also provides a foundation for the custom construction of pseudopotentials based on emerging functionals, for computationally-consistent modeling of alloys of growing complexity.

\subsection{Construction of disordered binary alloy supercells and structure optimization}
The special quasi-random structure (SQS) method was used to construct optimally-disordered supercells for modeling the six binaries. The SQS supercells were constructed using the MCSQS \cite{zunger1990special} module of the open source ATAT \cite{van2002alloy} software package. Structures with up to four-way correlations were considered, consistent with a range of $3\times$ the lattice constant ($\sim $~9--10 \text{\AA} for the binaries considered here). This range was determined based on our previous surface energy benchmarking studies for the bulk elements \cite{bijjala2025composition}, which showed that atomic interactions were not significant beyond $\approx$ 10 \text{\AA}. For all binaries and compositions, this procedure yielded 16-atom supercells with conventional BCC unit cells stacked in a $2\times2\times2$ fashion. Our construction is consistent with the previous work of Jiang {\it et al.} \cite{jiang2004first} on BCC MoNb, WTa, and CrFe, who performed extensive convergence studies of SQS supercell sizes for these binaries, and similarly concluded that 16-atom SQS’s are the optimal choice in terms of computational efficiency and accuracy of results. To test of the effect of SQS supercell size on our results, we performed 54-atom rigid-ion elastic constant calculations on $\rm{Mo_{0.5}Nb_{0.5}}$. A change of 0.8\% in $C_{11}$ and  changes of $<0.2\%$ in $C_{12}$ and $C_{44}$ were observed. These differences lie well within experimental errors, which can be as high as 10\% \cite{davies1977transducer,beg1976study}. The input SQS binary structures were constructed using lattice parameters obtained from the rule of mixtures applied to the constituent element lattice constants. Analogous to the procedure used for the bulk elements, the cell geometry (cell shape and volume) of each binary was optimized to converge the potential to $< 10^{-12}$ Ha. For the binaries, 50 $k$-points per \text{\AA} yielded $8\times 8\times 8$ special $k$-point grids. The Fermi surface of each system was integrated using Gaussian smearing \cite{verstraete2001smearing} with a width of 0.005 Ha. 10 valence bands/atom were used.

\subsection{Density functional perturbation theory}
Converged atomic structures obtained from geometry optimization calculations were used as input to the DFPT elastic tensor evaluations. $10 \times 10 \times 10$ $k$-point grids were used in the DFPT calculations. This ensured a clear separation between occupied and empty orbitals, allowing the application of DFPT to these binary metal alloys \cite{baroni2001phonons}. Abinit uses crystal symmetry to optimize computational efficiency. Since the structural relaxation of SQS supercells results in a loss of symmetry during unconstrained DFT relaxations, the Abinit symmetry checker was turned off in all calculations. DFPT calculations were carried out on relaxed structures in order to obtain derivatives of the occupied wave functions with respect to homogeneous strain perturbations $s_i$ (uniaxial and shear strains). The computed energy derivatives were used to compute the rigid-ion elastic tensors using the Analysis of Derivative DataBase ({\texttt{anaddb}) \cite{Gonze1997a, Gonze1997} module of Abinit. 

Computation of the nuclear relaxation contribution to the relaxed-ion elastic tensor as defined in Eq.~(\ref{eq:nuclear-relaxations})} requires determination of two intermediate quantities: the force-response internal strain tensor $\Lambda_{p\alpha, j}$ and the displacement-response internal strain tensor $\Gamma_{p\alpha, i}$, both associated with periodicity-preserving atomic displacements and homogenous strains ($s_i$, $s_j$,$s_k$,...). For a supercell with $N$ atoms, the atomic displacements are denoted by $u_{p\alpha}$, $u_{p\beta}$,..., with $\alpha$ and $\beta$ indexing Cartesian directions $x, y, z$, and the indices $p$ and $q$ labeling individual atoms.  The intermediate quantities are computed as \cite{hamann2005systematic}: 
\begin{align}
\label{eq:nuclear-contributions}
C_{ij}^{\text{nuclear}} = \frac{1}{\Omega_0} \sum_{p\alpha} \Gamma_{p\alpha, i} \Lambda_{p\alpha, j}
\end{align}
where
\begin{equation}
\Lambda_{p\alpha, j} = -\Omega_0 \frac{\partial^2 E}{\partial u_{p\alpha} \partial s_j}
\label{eq:Lambda}
\end{equation}
\begin{equation}
 \Gamma_{p\alpha, i} = \sum_{q\beta}\Lambda_{q\beta, i}{\left( K^{-1} \right)}_{p\alpha, q\beta} 
\label{eq:Gamma}
\end{equation}
and 
\begin{equation}
K_{p\alpha, q\beta} =  \Omega_0 \frac{\partial^2 E}{\partial u_{p\alpha} \partial u_{q\beta}}.
\end{equation}
Here $E$ is the energy per undeformed unit cell volume as defined above, and $\Omega_0$ is the initial volume of the crystal. Both $\Lambda_{nk}$ and $\Gamma_{nj}$ are computed in Abinit by performing response function calculations with respect to atomic displacements and homogeneous strains.

\noindent

\subsection{Elastic properties}
Within continuum elasticity theory, the elastic tensor of cubic structures \cite{zhao2007first}, including BCC crystals, have three independent elastic constants: $C_{11}$, $C_{12}$, and $C_{44}$. Since, as previously noted, full relaxation of the SQS structures breaks cubic symmetry, to enable comparison of DFPT results with experiment, the average elastic constants in the cubic structures were calculated using the Voigt-Hill approach \cite{hill1952elastic}:
\begin{align}
\label{eq:voigt-averages}
\bar{C}_{11} &= \frac{C_{11} + C_{22} + C_{33}}{3} \nonumber  \\
\bar{C}_{12} &= \frac{C_{12} + C_{23} + C_{31}}{3} \nonumber \\
\bar{C}_{44} &= \frac{C_{44} + C_{55} + C_{66}}{3}. 
\end{align}
To obtain composition-dependent values of these Voigt-averaged elastic constants, thermodynamic curve fitting was performed using a Redlich-Kister \cite{redlich1948algebraic} polynomial. Originally developed for fitting Gibbs free energies as a function of composition, it has also been used to fit elastic properties as a function of composition \cite{marker2018effects,korenev2022calculated,liu2010computational}:
\begin{align}
\label{eq:Redlich-Kister}
    \bar{C}_{ij}^{AB}(x_A) &= x_A\bar{C}_{ij}^{A} + (1-x_A)\bar{C}_{ij}^{B} \nonumber \\
                           &+ x_A(1-x_A)(I_0+I_1(2x_A-1)),
\end{align}
where $x_A$ is the atomic fraction of element A, and $\bar{C}_{ij}^{AB}(x_A)$ is the elastic constant of the AB binary with fractional composition $x_A$ of A. $\bar{C}_{ij}^{A}$ and $\bar{C}_{ij}^{B}$ are the elastic constants of pure A and B respectively, and $I_0$ and $I_1$ are interaction fitting parameters. Eq.~(\ref{eq:Redlich-Kister}) will be used to fit the elastic constant curves reported in Section~\ref{sec:Results} below. 

Based on the computed average elastic constants, the polycrystalline aggregate elastic moduli $B$ (bulk modulus), $G$ (shear modulus), $E$ (Young's modulus) and Possion's ratio \(\nu\) were obtained using the Voigt approximation \cite{simmons1971single}, which assumes a uniform strain in the crystal and gives an upper bound that correlates with the maximum value of a given property: 
\begin{equation}
\label{eq:B_voigt}
B_{\text{Voigt}} = \frac{\bar{C}_{11} + 2\bar{C}_{12}}{3} 
\end{equation}
\begin{align}
\label{eq:G_voigt}
G_{\text{Voigt}} &= \frac{\bar{C}_{11} - \bar{C}_{12} + 3\bar{C}_{44}}{5}
\end{align}
\begin{align} 
\label{eq:E_voigt}
E_{\text{Voigt}} &= \frac{9B_{\text{Voigt}}G_{\text{Voigt}}}{3B_{\text{Voigt}}+G_{\text{Voigt}}}
\end{align}
and 
\begin{align} 
\label{eq:v_voigt}
\nu_{\text{Voigt}} &= \frac{3B_{\text{Voigt}} - 2G_{\text{Voigt}}}{2(3B_{\text{Voigt}}+G_{\text{Voigt}})}.
\end{align}

The degree of elastic anisotropy for all binary alloys is characterized here using the shear $A_G$ and Young's $A_E$ anisotropy metrics \cite{nye1985physical, tromans2011elastic}:
\begin{align}
\label{eq:A_G}
    A_G &= \frac{S_{44}+S_{66}}{2S_{44}}\\
\label{eq:A_E}
    A_E&= \frac{S_{11}}{S_{33}},
\end{align}
where the $S_{ij}$ are the elements of compliance tensor, the inverse of the elastic tensor \cite{dieter1976mechanical}.

\subsection{Nuclear-relaxation fields}
If we decompose the intermediate quantities $\Gamma_{p\alpha, i}$ and $\Lambda_{p\alpha, j}$ (Eqs.~(\ref{eq:Lambda}) and (\ref{eq:Gamma})) used in computing the nuclear relaxation contribution to the relaxed-ion elastic tensor, into nuclear-relaxation contributions from each lattice site $\alpha$, Eq.~(\ref{eq:nuclear-contributions}) can be rewritten as:
\begin{align}
\label{eq:decomposition}
C_{ij}^{\text{nuclear}} = \frac{1}{\Omega_0} \sum_{p=1}^{N} \left (\sum_{\alpha} \Gamma_{p\alpha, i} \Lambda_{p\alpha, j}\right ),
\end{align}
where the inner index corresponds to summation over $(x,y,z)$ directions. This decomposition makes it possible to construct expressions for the nuclear-relaxation fields associated with each of the $B$, $G$, $E$, and $\nu$ elastic properties. First, each property is decomposed into rigid and nuclear relaxation contributions.  The latter are then expressed in terms of Voigt-averaged elastic properties and thus, in terms of elastic constants, which are in turn decomposed into contributions from each lattice site via Eq.~(\ref{eq:decomposition}).  The final representation in terms of nuclear relaxation atomic contributions is used to generate the nuclear relaxation field plots presented in Section~\ref{relaxation-fields}. The required decompositions into rigid and nuclear relaxation contributions are as follows.

For $B$ and $G$, the decomposition into rigid and nuclear relaxation contributions is exact.
For the bulk modulus $B$, we have:
\begin{align}
\label{eq:bulk_relaxed}
    B_{\text{relaxed}} = B_{\text{Voigt}} &= \frac{\bar{C}_{11} +2 \bar{C}_{12}}{3} \nonumber\\
    &= \frac{(\bar{C}_{11}^\text{{rigid}} - \bar{C}_{11}^\text{{nuclear}}) + 2(\bar{C}_{12}^\text{{rigid}} - \bar{C}_{12}^\text{{nuclear}}) }{3}\nonumber\\
     &= \frac{\bar{C}_{11}^\text{{rigid}} + 2\bar{C}_{12}^\text{{rigid}}}{3}
- \left (\frac{\bar{C}_{11}^\text{{nuclear}} + 2 \bar{C}_{12}^\text{{nuclear}} }{3} \right)\nonumber\\
&\equiv B_\text{{rigid}}-B_\text{{nuclear}}, 
\end{align}
with
\begin{equation}
    \label{eq:Bnuclear}
    B_\text{{nuclear}} \equiv \left (\frac{\bar{C}_{11}^\text{{nuclear}} + 2 \bar{C}_{12}^\text{{nuclear}} }{3} \right).
\end{equation}
Similarly, the decomposition of the shear modulus $G$ is given by:
\begin{align}
\label{eq:shear_relaxed}
    G_{\text{relaxed}} = G_{\text{Voigt}} &= \frac{\bar{C}_{11} - \bar{C}_{12} + 3\bar{C}_{44}}{5} \nonumber\\
    &= \frac{(\bar{C}_{11}^\text{{rigid}} - \bar{C}_{11}^\text{{nuclear}}) - (\bar{C}_{12}^\text{{rigid}} - \bar{C}_{12}^\text{{nuclear}}) +3(\bar{C}_{44}^\text{{rigid}} - \bar{C}_{44}^\text{{nuclear}})}{5}\nonumber\\
     &= \frac{\bar{C}_{11}^\text{{rigid}} - \bar{C}_{12}^\text{{rigid}}  +3 \bar{C}_{44}^\text{{rigid}}}{5}
- \left (\frac{\bar{C}_{11}^\text{{nuclear}} - \bar{C}_{12}^\text{{nuclear}}  +3 \bar{C}_{44}^\text{{nuclear}}}{5} \right)\nonumber\\
&\equiv G_\text{{rigid}}-G_\text{{nuclear}},
\end{align}
with 
\begin{equation}
    \label{eq:G-nuclear}
G_\text{{nuclear}} \equiv \left (\frac{\bar{C}_{11}^\text{{nuclear}} - \bar{C}_{12}^\text{{nuclear}}  +3 \bar{C}_{44}^\text{{nuclear}}}{5} \right).
\end{equation}
In these expressions, by analogy with Eq.~(\ref{eq:voigt-averages}), we have defined
\begin{equation}
    \bar{C}_{11}^\kappa \equiv \frac{{C}_{11}^\kappa+{C}_{22}^\kappa+{C}_{33}^\kappa}{3}
\end{equation}
\begin{equation}
    \bar{C}_{12}^{\kappa} \equiv \frac{{C}_{12}^{\kappa}+{C}_{23}^{\kappa}+{C}_{31}^{\kappa}}{3}
\end{equation}
\begin{equation}
    \bar{C}_{44}^{\kappa} \equiv \frac{{C}_{44}^{\kappa}+{C}_{55}^{\kappa}+{C}_{66}^{\kappa}}{3},
\end{equation}
with $\kappa$ = `rigid' or `nuclear'.

For the Young's modulus $E$ and Possion's ratio $\nu$, the decompositions into rigid and nuclear components are no longer exact, due to the presence of the combination $3B_{\text{Voigt}} + G_{\text{Voigt}}$ in the denominators of Eqs.~(\ref{eq:E_voigt}) and (\ref{eq:v_voigt}). However, it is possible to derive approximate expressions for the decompositions by noting that 
\begin{equation}
\label{eq:small}
    \frac{3B_n + G_n}{3B_r + G_r} \ll 1, 
\end{equation}
where we have abbreviated $B_n = B_{\text{nuclear}}$, $G_n = G_{\text{nuclear}}$, $B_r = B_{\text{rigid}}$, and $G_r = G_{\text{rigid}}$. The relevant derivations are given in Appendix A.1. For the Young's modulus $E$, we have: 
\begin{align}
 \label{eq:Young's_relaxed}
E_{\text{relaxed}} = E_{\text{Voigt}} &= E_\text{{rigid}}-E_\text{nuclear},
\end{align}
with 
\begin{equation}
E_\text{rigid} = \frac{9B_{\text{rigid}}G_{\text{rigid}}}{3B_{\text{rigid}}+G_{\text{rigid}}}
\end{equation} 
and
\begin{equation}
E_\text{nuclear} \approx E_\text{rigid} \left(E_2 - E_1 + E_2E_1 \right ).
\label{eq:E_nuclear}
\end{equation}
$E_1$ and $E_2$ are defined in Appendix~A.1 as
\begin{equation}
    E_1 \equiv \frac{3B_n+G_n}{3B_r+G_r}   
\end{equation}
and 
\begin{equation}
    E_2 \equiv \frac{B_r G_n + B_n G_r - B_n G_n}{B_r G_r}.
\end{equation}

Similarly, the decomposition of Poisson's ratio $\nu$ is given by: 
\begin{align}
 \label{eq:Possion's_relaxed}
\nu_{\text{relaxed}} = \nu_{\text{Voigt}} &= \nu_\text{{rigid}}-\nu_\text{nuclear},
\end{align}
where 
\begin{equation}
\nu_\text{rigid} = \frac{3B_{\text{rigid}}-2G_{\text{rigid}}}{2\left(3B_{\text{rigid}}+G_{\text{rigid}}\right)}
\end{equation}
and
\begin{equation}
\nu_\text{nuclear} \approx\nu_{\text{rigid}} \left (\nu_2 - \nu_1 + \nu_2\nu_1\right ).
\label{eq:nu_nuclear}
\end{equation}
$\nu_1$ and $\nu_2$ are defined in Appendix~A.1 as
\begin{equation}
    \nu_1 = \frac{3B_n + G_n}{3B_r + G_r}
\end{equation}
and 
\begin{equation}
    \nu_2 = \frac{3B_n - 2G_n}{3B_r - 2G_r}.
\end{equation}

Note that although $\nu_1 = E_1$, for clarity, separate symbols are used in the final expressions for $E_\text{nuclear}$ and $\nu_\text{nuclear}.$ To test (and confirm) the $E_1 << 1$ assumption, the values of $E_1$ were tabulated for all six binary alloys across all compositions.  In addition, as a check of the new equations and associated analysis code for the relaxation field decompositions of $E$ and $\nu$, the Abinit-computed nuclear relaxations were compared with the sum of approximated atomic nuclear relaxation values; excellent agreement was obtained. The results of both tests are summarized in Tables 3 and 4, Appendix A.2.

The decompositions of the elastic constants and elastic properties into lattice-dependent nuclear relaxation contributions will be used in the relaxation field analysis of Section~\ref{relaxation-fields}.

\section{Results}
\label{sec:Results} 
\subsection{Rigid-ion vs.~relaxed-ion elastic constants}
The rigid-ion and relaxed-ion tensors were computed for the binary alloys as described in Section~\ref{sec:CompMeth}. Prior to proceeding with further analysis, we confirmed that the results for both the rigid-ion and relaxed-ion elastic tensors satisfied the well-known Born stability criteria \cite{born1940stability}:
\begin{align}
    C_{11} - C_{12} &> 0;\ \ C_{11} +2 C_{12} > 0;\  \ C_{44}  > 0 
\end{align}
and condition for mechanical stability of solids under zero stress \cite{born1954dynamical, kittel2018introduction}:
\begin{align}
    C_{11}^2 - C_{12}^2 &> 0, C_{11} > 0 \nonumber\\
    \boldsymbol{\lambda} &> 0,
\end{align}
where $\boldsymbol{\lambda}$ is the vector of six eigenvalues of the $6 \times 6$ elastic tensor. 

The three independent Voigt-averaged elastic constants, $\bar{C}_{11}$, $\bar{C}_{12}$ and $\bar{C}_{44}$, were computed together with their standard deviation values, for each of the BCC binary alloys starting from their rigid-ion and relaxed-ion elastic tensors. The results are listed in Table~\ref{tab:rigid_relaxed} for three different compositions per AB alloy: A = 25 at.\%, 50 at.\%, and 75 at.\%. The standard deviation $\sigma_{C_{11}}$ was computed as:
\begin{equation}
    \label{eq:std}
    \sigma_{C_{11}} = {\frac{1}{3}[(\bar{C}_{11}-C_{11})^2 + (\bar{C}_{11}-C_{22})^2 + (\bar{C}_{11}-C_{33})^2]}^{1/2},
\end{equation}
and the standard deviations for $\bar{C}_{12}$ and $\bar{C}_{44}$ were computed similarly. The standard deviation values are observed to be non-zero for both the rigid-ion and relaxed-ion elastic tensors in the binary alloys of A = 25 at.\% and 75 at.\%, and zero in the binary alloys of A = 50 at.\% for all $\bar{C}_{11}$, $\bar{C}_{12}$ and $\bar{C}_{44}$. The standard deviations are of similar magnitudes for both the rigid-ion and relaxed-ion methods, with no particular trend observed among the binary alloys. 

The differences between the rigid-ion and relaxed-ion values in Table~\ref{tab:rigid_relaxed} correspond to relaxation values (Eq.~(\ref{eq:nuclear-relaxations})). For $\bar{C}_{11}$, relaxation values for the same-group binary alloys MoW and NbTa are close to 0 at all compositions. However, in the binary alloys MoNb, MoTa, WNb and WTa, whose constituent elements come from different groups (V and VI) of the Periodic Table, relaxation values are maximized at A = 50 at.\%, with the ordering $({\rm WTa} \approx {\rm WNb}) > ({\rm MoTa} \approx{\rm MoNb})$. 

For $\bar{C}_{12}$, WNb exhibits the greatest relaxation value at all compositions, with maximum value at A = 50 at.\%. A similar trend to that of $\bar{C}_{11}$ is observed in the same-group binary alloys. However, in different-group binary alloys, MoTa exhibits negative relaxations at all compositions, with a maximum magnitude of nearly 1 GPa at A = 50 at.\% and values close to 0 at A = 25 at.\% and 75 at.\%.  This is notable, since a negative relaxation value corresponds to a greater relaxed-ion elastic constant value compared to the rigid-ion elastic constant value---i.e., it reflects improvement in the elastic constant property with alloying.  Unlike for $\bar{C}_{11}$, no common trend in $\bar{C}_{12}$ is observed in different-group binary alloys.  

In the case of $\bar{C}_{44}$, there is a positive relaxation value for all compositions, but no common trend is observed for the same-group binary alloys. The relaxation values for MoW are near zero for all compositions, while the values for NbTa are larger and positive across all compositions, with a peak value at A = 50 at.\%. Among different group binary alloys, WNb shows positive but near-zero values for all compositions, while MoNb, MoTa and WTa, show a finite positive relaxation value. A common trend of peak nuclear relaxation value at  A = 25 at.\%  followed by A = 50 at.\% and A = 75 at.\% is observed for three of the different-group binary alloys, with magnitude ordering ${\rm MoTa} > {\rm WTa} > {\rm MoNb}$. 
\begin{table*}[!ht]   
    \caption{Composition-dependent Voigt-averaged rigid-ion and relaxed-ion elastic constants for the binary alloys (Eq.~(\ref{eq:voigt-averages}), with standard deviations ($\pm$).)}
    \centering
    \scalebox{0.82}{
    \begin{tabular}{|c|c|c|c|c|c|c|c|}
        \hline
        \multicolumn{2}{|c|}{\textbf{Composition}} & \textbf{$\bar{C}_{11}$} (GPa) & \textbf{$\bar{C}_{12}$} (GPa)& \textbf{$\bar{C}_{44}$} (GPa)& \textbf{$\bar{C}_{11}$}(GPa) & \textbf{$\bar{C}_{12}$}(GPa) & \textbf{$\bar{C}_{44}$}(GPa) \\
        \hline
        AB & at.\% A & \multicolumn{3}{|c|}{\textbf{Rigid-ion}}& \multicolumn{3}{|c|}{\textbf{Relaxed-ion}}\\
        \hline
        \multirow{3}{*}{MoNb} &  25  & $297.27 \pm 5.5$   & $136.83\pm1.5$   & $33.07\pm0.7$
  & $294.62\pm7.26$ & $136.37\pm1.63$ & $32.05\pm0.83$ \\
        &50 & $358.70\pm 0.00$   & $138.91\pm 0.00$   & $64.91\pm 0.00$   & $355.59\pm 0.00$   & $138.60\pm 0.00$ & $64.53\pm 0.00$ \\
        &75 & $423.88\pm9.67$   & $145.49\pm0.55$   & $84.05\pm1.43$   & $422.56\pm8.69$   & $145.33\pm0.53$ & $83.75\pm1.38$ \\
        \hline
        \multirow{3}{*}{MoTa} &  25  & $302.24\pm1.66$   & $162.35\pm2.07$   & $60.59\pm0.96$
  & $299.52\pm2.50$ & $162.19\pm1.97$ & $55.29\pm0.93$ \\
        &50 & $373.72\pm0.00$   & $159.89\pm0.00$   & $85.31\pm0.00$   & $369.97\pm0.00$   & $160.64\pm0.00$ & $83.17\pm0.00$ \\
        &75 & $429.57\pm3.37$   & $155.30\pm0.90$   & $92.48\pm0.77$   & $428.28\pm3.28$   & $155.54\pm0.91$ & $92.02\pm0.76$ \\
        \hline
        \multirow{3}{*}{MoW} &  25  & $499.36\pm1.18$   & $188.39\pm0.15$   & $130.25\pm0.31$
  & $499.13\pm1.13$ & $188.17\pm0.14$ & $130.12\pm0.28$ \\
        &50 & $486.82\pm0.00$   & $180.13\pm0.00$   & $121.21\pm0.00$   & $486.38\pm0.00$   & $179.71\pm0.00$ & $120.97\pm0.00$ \\
        &75 & $467.32\pm1.02$   & $168.59\pm0.28$   & $109.41\pm0.42$   & $467.11\pm1.02$   & $168.39\pm0.28$ & $109.31\pm0.40$ \\
        \hline    
        \multirow{3}{*}{NbTa} &  25  & $256.28\pm0.66$   & $152.21\pm0.43$   & $58.39\pm0.23$
  & $256.20\pm0.67$ & $152.17\pm0.40$ & $57.20\pm0.68$ \\
        &50 & $250.88\pm0.00$   & $144.79\pm0.00$   & $44.68\pm0.00$   & $250.69\pm0.00$   & $144.68\pm0.00$ & $41.85\pm0.00$ \\
        &75 & $245.81\pm3.11$   & $137.39\pm0.45$   & $30.64\pm0.27$   & $245.70\pm3.07$   & $137.32\pm0.42$ & $28.98\pm0.85$ \\
        \hline
        \multirow{3}{*}{WNb} &  25  & $300.06\pm3.73$   & $149.68\pm1.19$   & $37.30\pm1.06$
  & $296.69\pm4.22$ & $147.89\pm1.63$ & $37.17\pm1.03$ \\
        &50 & $377.12\pm0.00$   & $157.79\pm0.00$   & $72.70\pm0.00$   & $371.88\pm0.00$   & $155.53\pm0.00$ & $72.46\pm0.00$ \\
        &75 & $450.72\pm0.56$   & $174.74\pm0.34$   & $108.10\pm0.66$   & $448.83\pm1.20$   & $173.74\pm0.53$ & $107.77\pm0.56$ \\
        \hline
        \multirow{3}{*}{WTa} &  25  & $307.83\pm1.02$   & $173.17\pm1.97$   & $70.16\pm1.15$
  & $304.84\pm1.22$ & $171.96\pm2.26$ & $67.91\pm0.46$ \\
        &50 & $377.61\pm0.00$   & $180.22\pm0.00$   & $89.32\pm0.00$   & $372.38\pm0.00$   & $179.41\pm0.00$ & $88.02\pm0.00$ \\
        &75 & $456.06\pm0.62$   & $184.67\pm0.47$   & $116.53\pm0.86$   & $454.37\pm1.93$   & $184.27\pm0.51$ & $116.26\pm0.84$ \\
        \hline 
    \end{tabular}}
    \label{tab:rigid_relaxed}
\end{table*} 

\subsection{Rigid-ion elastic constants}
Despite the lattice distortions in the binary alloys, the magnitudes of the relaxations lie within the range of the standard deviations and within 2\% of the rigid-ion elastic constant values. These differences lie within experimental errors, which can be as high as 10\% \cite{davies1977transducer,beg1976study}, depending on the experimental conditions and technique employed. Hence, the computationally faster rigid-ion elastic tensor calculations were performed for the additional compositions of  A = 12.5 at.\%, 37.5 at.\%, 62.5 at.\% and 87.5 at.\% for all of the AB binary alloys. This allowed us to capture the effects of alloying on the elastic constants of BCC refractory binary alloys for an extended set of compositions. 
\begin{figure*}[htbp]
\centering
\includegraphics[width=1\linewidth]{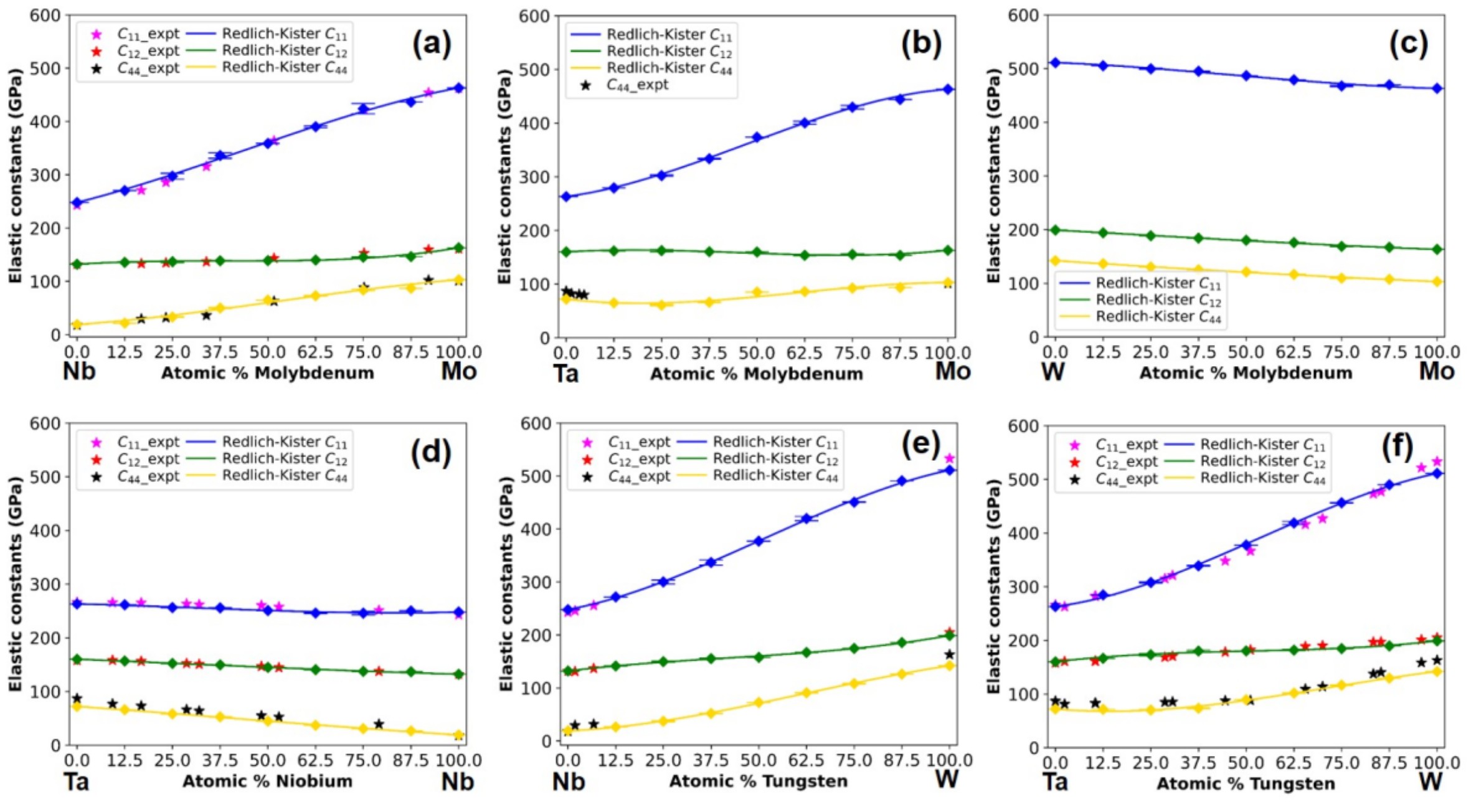}
\caption{\label{fig:rigid-ET} Composition-dependent Voigt-averaged rigid-ion elastic constants $\bar{C}_{11}^{rigid}$, $\bar{C}_{12}^{rigid}$ and $\bar{C}_{44}^{rigid}$ (solid lines, Redlich-Kister polynomial fit to computed values (filled diamonds)), compared with experimental elastic constants (stars) for (a) MoNb \cite{hubbell1972elastic}; (b) MoTa \cite{armstrong1970influence}; (c) MoW; (d) NbTa \cite{fisher1981effects}; (e) WNb \cite{frey1978elastic}; and (f) WTa \cite{anderson1980single}. Computed error bars correspond to standard deviation with respect to Voigt averages (Eq.~(\ref{eq:voigt-averages})).}
\end{figure*}

Fig.~\ref{fig:rigid-ET} compares the calculated rigid-ion elastic constants ($\bar{C}_{11}^{rigid},$ $\bar{C}_{12}^{rigid},$ and $\bar{C}_{44}^{rigid}$) for all compositions (filled diamonds) with experimental values (stars) for the binary alloys. The rigid-ion elastic constants are fitted to the Redlich-Kister polynomial given in Eq.~(\ref{eq:Redlich-Kister}) (solid line).  For the same-group binary alloys (Figs.~\ref{fig:rigid-ET}(c) and (d)), elastic constants are observed to vary linearly, similar to the empirical rule of Vegard's law \cite{denton1991vegard}, while nonlinear behavior is observed for the different-group binary alloys. For the different-group binary alloys, the Ta-containing binary alloys (Figs.~\ref{fig:rigid-ET}(b) and (f)) exhibit a greater degree of nonlinearity compared to the Nb-containing binary alloys (Figs.~\ref{fig:rigid-ET}(a) and \ref{fig:rigid-ET}(e)). For  $\bar{C}_{11}^{rigid}$ and $\bar{C}_{44}^{rigid}$ in the different-group binary alloys, a negative deviation from Vegard's law is observed for the compositions with ${\rm X} > 50$ at.\% where X is the element with lower elastic modulus value of the constituent elements. However, for $\bar{C}_{12}^{rigid}$, 
the trend is reversed, {\it i.e.}, a positive deviation from Vegard's law is observed for the compositions when ${\rm X} > 50$ at.\% where X is the element with lower elastic modulus value of the constituent elements.
\begin{figure*}[htbp]
\centering
\includegraphics[width=1\linewidth]{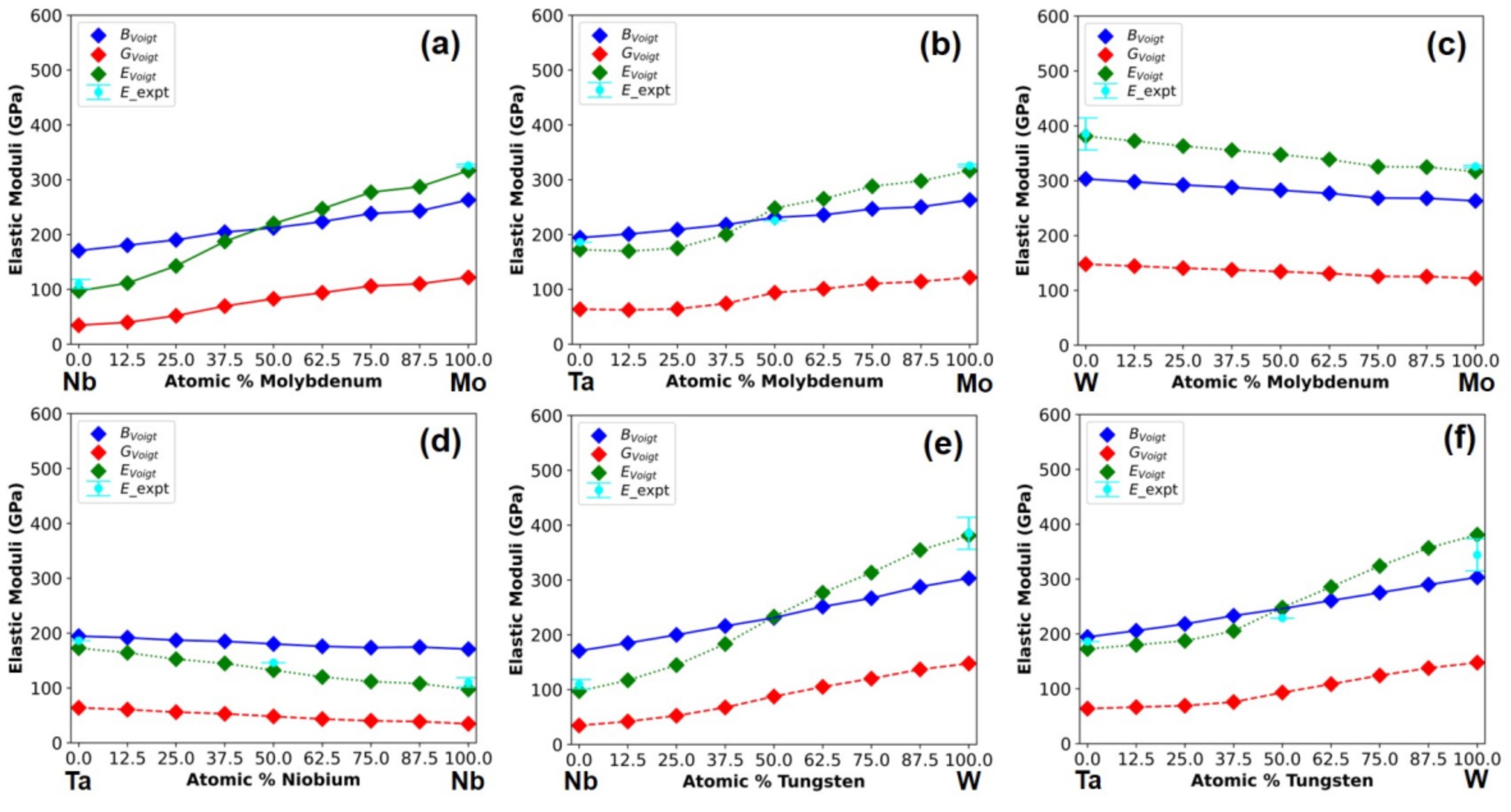}
\caption{\label{fig:Voight-moduli} Composition-dependent elastic moduli $B$, $G$ and $E$ computed using the Voigt method (Eqs. (\ref{eq:B_voigt})--(\ref{eq:E_voigt})), as a function of composition, compared with experimental $E$ (circles) for (a) MoNb \cite{mccoy1964mechanical, campbell2008elements, WebElements-website}; (b) MoTa \cite{mccoy1964mechanical, campbell2008elements, WebElements-website, khakurel2021machine}; (c) MoW \cite{mccoy1964mechanical, campbell2008elements, WebElements-website}; (d) NbTa \cite{mccoy1964mechanical, campbell2008elements, WebElements-website, khakurel2021machine}; (e) WNb \cite{mccoy1964mechanical, campbell2008elements, WebElements-website}; and (f) WTa \cite{mccoy1964mechanical, campbell2008elements, WebElements-website, khakurel2021machine}. Error bars correspond to the standard deviation about averaged experimental values.}
\end{figure*}

\subsubsection{Comparison with experiment}
Experimental values at room temperature for all elastic constants were obtained from the literature as measured using high-frequency ultrasonic methods. For MoNb \cite{hubbell1972elastic},  NbTa \cite{fisher1981effects} and WNb \cite{frey1978elastic}, elastic constants were obtained by measuring sound velocities using the pulse superposition method, and for MoTa elastic constants were obtained by measuring sound velocities using the thin rod resonance technique. High-frequency methods are generally considered to have higher accuracy than static or low frequency methods \cite{armstrong1970influence}. Excellent agreement between the computed rigid-ion elastic coefficients and available experimental data is observed, confirming the accuracy of the rigid-ion method for computing the elastic tensors in the present class of binary alloys.

It should be noted that our electronic structure calculations are performed at 0~K, while the experimental results are obtained at 298~K; however, refractory metals/alloys are known for the stability of their mechanical properties in this temperature range \cite{hubbell1972elastic,armstrong1970influence,frey1978elastic,anderson1980single}. Experimental data for the temperature dependence of the elastic constants $C_{11}$, $C_{12}$ and $C_{44}$ is available for the bulk elements Mo, Ta, and W \cite{featherston1963elastic}, Nb \cite{hubbell1972elastic}, and---as a function of composition---for the binary alloys MoNb \cite{hubbell1972elastic} and WTa \cite{anderson1980single}. The values of all  elastic constants decrease monotonically as a function of temperature except for $C_{44}$ of MoNb, which increases with temperature for several compositions \cite{hubbell1972elastic}.  For the bulk elements, the differences between effective 0~K (measured 4.2~K) and 298~K $C_{11}$ and $C_{12}$ values are all $< 2.2\%$.  For $C_{44}$, the differences between the 4.2~K and 298~K values range from 1.5\% for W to 6.3\% for Ta. For MoNb, the differences between measured values at 83~K and 298~K for both $C_{11}$ and $C_{12}$ are $< 3\%$ for all reported compositions \cite{hubbell1972elastic}. For $C_{44}$, the differences in values between 83~K and 298~K range from $< 1\%$ to 13.9\% (where the magnitude of $C_{44}$ is small) depending on composition. For WTa, the differences between measured values at 83~K and 298~K for both $C_{11}$ and $C_{12}$ are $< 2\%$ for all reported compositions \cite{anderson1980single}. For $C_{44}$, the differences in values between 83~K and 298~K range from $< 1\%$ to 6.5\% depending on composition. Overall, these temperature variations lie within expected experimental error bars of $\approx$ 10 \% for elastic constant measurements \cite{davies1977transducer,beg1976study}.

\subsection{Derived mechanical properties}
Polycrystalline elastic moduli were derived from the set of computed, Voigt-averaged rigid-ion elastic constants ($\bar{C}_{11}^{rigid}$, $\bar{C}_{12}^{rigid}$ and $\bar{C}_{44}^{rigid}$) for the binary alloys as a function of composition. The Voigt method \cite{simmons1971single} was used to calculate the polycrystalline elastic moduli. Fig.~\ref{fig:Voight-moduli} summarizes $B$, $G$ and $E$ for all refractory binary alloys along with experimental values of $E$ \cite{mccoy1964mechanical, campbell2008elements, WebElements-website, khakurel2021machine}. Similar to the behavior observed for the elastic constants, same-group binary alloys exhibit a linear trend as a function of composition (Figs.~\ref{fig:Voight-moduli}(c) and \ref{fig:Voight-moduli}(d)) for all elastic moduli. While $E > B$ for all MoW compositions, $E < B$ in NbTa across all compositions. Among different-group binary alloys, the bulk modulus is observed to vary linearly with composition, while $G$ and $E$ display nonlinear behavior. Similar to $\bar{C}_{11}^{rigid}$, $\bar{C}_{12}^{rigid}$ and $\bar{C}_{44}^{rigid}$, the deviation from linearity for $G$ and $E$ is greater in case of the Ta-containing binary alloys (Figs.~\ref{fig:Voight-moduli}(b) and (f)) than for the Nb-containing binary alloys (Figs.~\ref{fig:Voight-moduli}(a) and (e)).
\begin{figure*}[htbp]
\centering
\includegraphics[width=1\linewidth]{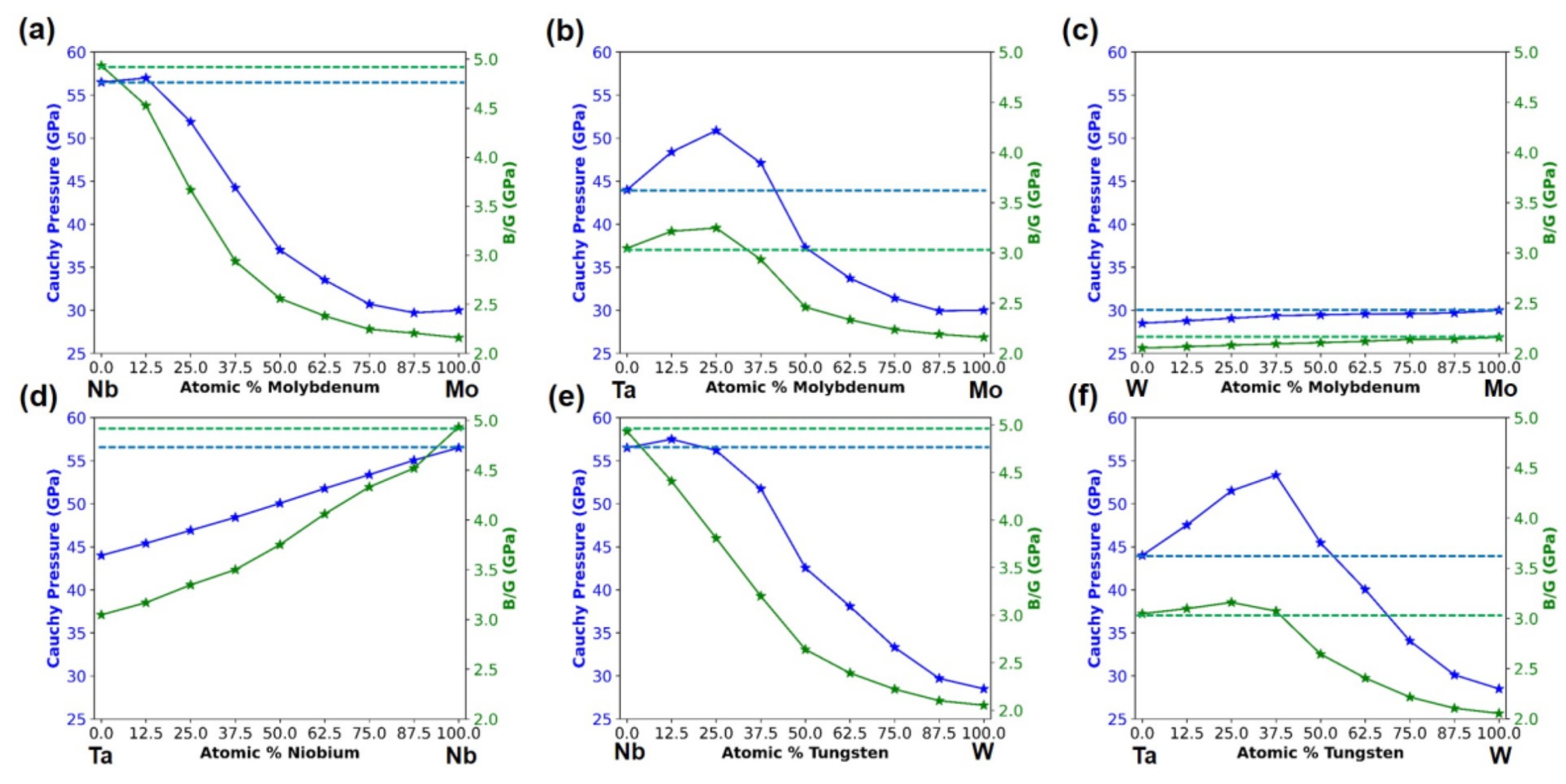}
\caption{\label{fig:Pugh-ratio} Composition-dependent Pugh's ratio ($B/G$; green) and Cauchy pressure (($C_{12} - C_{44}$); blue) for (a) MoNb; (b) MoTa; (c) MoW; (d) NbTa; (e) WNb; and (f) WTa. Horizontal lines indicate maximum Cauchy pressure or Pugh's ratio of the constituent elements.}
\end{figure*}
Fig.~\ref{fig:Pugh-ratio} shows Pugh's ratio ($B/G$) and the Cauchy pressure ($(C_{12}-C_{44})/2$) for the binary alloys as a function of composition.
As noted above, Pugh's ratio $B/G$ \cite{pugh1954xcii} and the Cauchy pressure \cite{pettifor1992theoretical, pettifor1991bonding} are widely-used to anticipate the ductility of an alloy. A material is expected to exhibit brittle behavior if Pugh's ratio is less than 2 \cite{gilman2003electronic, niu2012extra, thompson2018predicting}, and is intrinsically ductile if the Cauchy pressure is positive \cite{pettifor1992theoretical, pettifor1991bonding}. Similar to the behavior observed for the elastic constants and elastic moduli, MoW and NbTa show a linear trend as a function of composition for both Pugh's ratio and the Cauchy pressure (Figs.~\ref{fig:Pugh-ratio}(c) and \ref{fig:Pugh-ratio}(d)). Among different-group binary alloys, Nb-containing binary alloys displayed different behavior for Pugh's ratio and the Cauchy pressure as a function of composition, compared to the Ta-containing binary alloys. In the case of the Nb-containing binary alloys MoNb and WNb, the $B/G$ value is always less than that of the constituent elements across the entire composition range ((Figs.~\ref{fig:Pugh-ratio}(a) and \ref{fig:Pugh-ratio}(e)). However, the Cauchy pressures in MoNb and WNb have higher values than those of the constituent elements at 12.5 at.\% Nb composition in both binary alloys. In the case of the Ta-containing binary alloys MoTa and WTa, both $B/G$ and the Cauchy pressure values are observed to be greater than the values for their constituent elements for some compositions (Figs.~\ref{fig:Pugh-ratio}(b) and \ref{fig:Pugh-ratio}(f)). In MoTa, the $B/G$ and Cauchy pressure values are greater than that of the ductile constituent element Ta up to $\approx 35$ at.\% and $\approx40$ at.\% of Ta, respectively. In WTa, the $B/G$ and Cauchy pressure values are greater than their ductile constituent element Ta up to $\approx37.5$ and $\approx52.5$ at.\% of Ta, respectively. 

\subsection{Elastic anisotropy and elastic heterogeneity as a function of composition}
The finite relaxations and standard deviation values in the elastic constants for the binary alloys summarized in Table~\ref{tab:rigid_relaxed} reflect different degrees of spatial variations (elastic heterogeneity) in these materials. In order to quantitate the differences in elastic anisotropies among the binaries, at different compositions, we computed the $A_G$ and $A_E$ metrics defined in Eqs.~(\ref{eq:A_G}) and (\ref{eq:A_E}). Table ~\ref{tab:anisotropy} compares the calculated anisotropies for the relaxed-ion elastic constants. In the case of an isotropic bulk material, these values will be unity, and the deviation from unity corresponds to the degree of anisotropy in the material. The anisotropy measures $A_G$ and $A_E$ are close to one in all binary alloys, corresponding to near-isotropic behavior. These anisotropies reflect the local anisotropy at the atomic level, corresponding to different atomic bond strengths in different crystallographic planes. Although the values of $A_G$ and $A_E$ suggest that the binary alloys are near-isotropic, these measures do not provide insight into the spatial variations of the elastic moduli. Since rigid-ion elastic constants are uniform across the lattice and relaxation values are added as a correction to obtain relaxed-ion elastic constants, we therefore define two new metrics, $R_G$ and $R_E$, expressed in terms of the respective ratios of the nuclear and relaxed contributions. An elastic heterogeneity constant with a value of unity indicates the absence of spatial variation in elastic properties. These characterize the average changes in the shear modulus and Young's modulus  values at different atomic sites across the lattice:

\begin{equation}
  \label{eq:R_G}
    {R_G} = 1-\frac{G_{\text{nuclear}}}{G_{\text{relaxed}}}
\end{equation}
and
\begin{equation}
  \label{eq:R_E}
    R_E = 1-\frac{E_{\text{nuclear}}}{E_{\text{relaxed}}},
\end{equation}
where $G_{\text{nuclear}}$ and $G_{\text{relaxed}}$ are defined in Eqs.~(\ref{eq:shear_relaxed}) and (\ref{eq:G-nuclear}) and $E_{\text{nuclear}}$ and $E_{\text{relaxed}}$ are defined in Eqs.~(\ref{eq:Young's_relaxed}) and (\ref{eq:E_nuclear}). The values of $R_G$ and $R_E$ are also listed in Table~\ref{tab:anisotropy}. 
\begin{table*}[!ht]
    \centering
    \caption{Comparison of $A_G$ (Eq.~({\ref{eq:A_G}})) with $R_G$ (Eq.~({\ref{eq:R_G}})) and $A_E$ (Eq.~({\ref{eq:A_E}}))  with $R_E$ (Eq.~({\ref{eq:R_E}})) and  for all binary alloys and different compositions.}
    \scalebox{0.85}{
    \begin{tabular}{|c|c|c|c|c|c|c|c|c|c|c|c|}
        \hline
        \multicolumn{2}{|c|}{\textbf{Composition}} & \multirow{2}{*}{$A_G$} & \multirow{2}{*}{$R_G$} & \multirow{2}{*}{$A_E$}& \multirow{2}{*}{$R_E$}&
        \multicolumn{2}{|c|}{\textbf{Composition}} & \multirow{2}{*}{$A_G$} & \multirow{2}{*}{$R_G$} & \multirow{2}{*}{$A_E$}& \multirow{2}{*}{$R_E$} \\
        \cline{1-2} \cline{7-8}
        AB & at.\% of A                            &                        &                        &                        & &
        AB & at.\% of A                            &                        &                        &   &                     \\
        \hline
        \multirow{3}{*}{MoNb} &  25  & 1.032&0.979&1.049&0.981 & \multirow{3}{*}{MoTa} &  25  & 1.007&0.939&0.987&0.945  \\
        &50 & 1.000&0.991&1.000&0.991& &50 & 1.000&0.977&1.000&0.979 \\
        &75 & 0.984&0.996&0.959&0.996& &75 & 0.996&0.995&1.015&0.995  \\
        \hline

        \multirow{3}{*}{MoW} &  25  & 1.000&0.999&1.000&0.999 & \multirow{3}{*}{NbTa} &  25  & 1.009&0.986&1.003&0.988 \\
        &50 & 1.000&0.999&1.000&0.999& &50 & 1.000&0.961&1.000&0.966 \\
        &75 & 1.001&1.000&1.003&0.999& &75 & 1.013&0.973&1.004&0.976 \\
        \hline
        \multirow{3}{*}{WNb} &  25  & 0.982&0.992&0.970&0.992 & \multirow{3}{*}{WTa} &  25  & 0.999&0.975&1.000&0.976 \\
        &50 & 1.000&0.992&1.000&0.991& &50 & 1.000&0.982&1.000&0.983 \\
        &75 & 0.997&0.997&1.004&0.997& &75 & 0.995&0.994&1.004&0.997 \\
        \hline
    \end{tabular}}
    \label{tab:anisotropy}
\end{table*}

\begin{figure*}[htbp]
\centering
\includegraphics[width=1.0\linewidth]{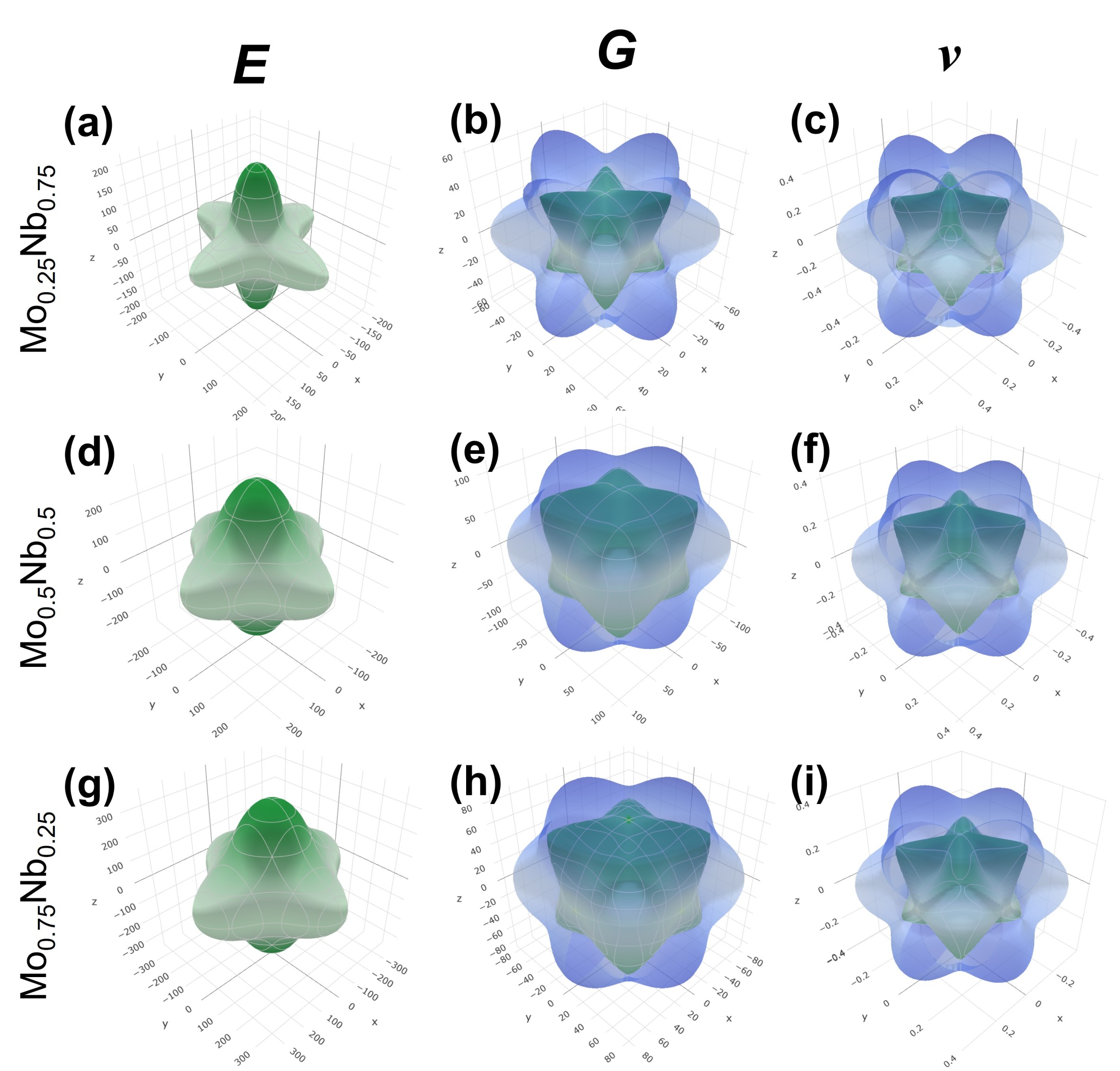}
\caption{\label{fig:anisotropy_MoNb} Anisotropy plots of $G$ (GPa), $E$ (GPa), and $\nu$ for: (a)-(c)  $\rm{Mo_{0.25}Nb_{0.75}}$;\, (d)-(f)  $\rm{Mo_{0.5}Nb_{0.5}}$; and (g)-(h) $\rm{Mo_{0.75}Nb_{0.25}}$.  Plots generated using the ELATE tool \cite{gaillac2016elate}. The maximum and minimum values of $G$, $E$ and $\nu$ are represented as transparent blue and solid green surfaces respectively.}
\end{figure*}

To further analyze the extent of elastic tensor property anisotropy in the refractory binary alloys, 3D plots of the orientation-dependent $G$, $E$ and $\nu$, for three compositions each, were generated using the ELATE tool \cite{gaillac2016elate}.  The results for MoNb and WTa are shown in Figs.~\ref{fig:anisotropy_MoNb} and \ref{fig:anisotropy_WTa}; the results for MoTa, MoW, NbTa, and WNb can be found in the Appendix.  The axes of the 3D plots in the figures correspond to the crystal orientations $[1 \, 0 \, 0]$, $[0 \,  1\, 0]$, $[0\, 0 \, 1]$ in the $x$, $y$, and $z$ directions, respectively. The maximum and minimum values of $G$ and $\nu$ are represented as transparent blue and solid green surfaces. The extent of anisotropy is seen to be greatest for MoNb.  For the Young's modulus $E$, $[ 1 \, 0 \, 0 ]$ is the stiffest direction and $[ 1 \, 1 \, 1 ]$ is the most compliant direction in all compositions of the binary alloys except for $\rm{Nb_{0.25}Ta_{0.75}}$ and $\rm{W_{0.25}Ta_{0.75}}$, which have nearly perfect spherical shapes, corresponding to elastic isotropy. $\rm{Mo_{0.25}Nb_{0.75}}$, $\rm{W_{0.25}Nb_{0.75}}$ and $\rm{Nb_{0.75}Ta_{0.25}}$ deviate strongly away from spherical, indicating high elastic anisotropy. Interestingly, WTa exhibits near-isotropic behavior at all compositions as seen in Fig.~\ref{fig:anisotropy_WTa}. 
Similar trends are observed for $G$ and $\nu$.  The patterns observed in the 3D plots are consistent with the reduced metrics $A_E$ and $A_G$ in Table~\ref{tab:anisotropy}. For $A_E$, all binaries showed near-isotropic behavior with deviations away from 1 of $\le 1\%$ except for $\rm{Mo_{0.25}Nb_{0.75}}$, $\rm{Mo_{0.75}Nb_{0.25}}$ and $\rm{W_{0.25}Nb_{0.75}}$, where the deviations are $> 3\%$. For $A_G$, all binaries exhibited near-isotropic behavior (deviations of $< 1\%$), except for $\rm{Mo_{0.25}Nb_{0.75}}$, $\rm{Mo_{0.75}Nb_{0.25}}$, $\rm{Nb_{0.75}Ta_{0.25}}$ and $\rm{W_{0.25}Nb_{0.75}}$, which exhibit greater deviations away from unity. For both $R_G$ and $R_E$, all binaries showed near-isotropic behavior (deviations of $< 1\%$), except for $\rm{Mo_{0.25}Nb_{0.75}}$, $\rm{Mo_{0.25}Ta_{0.75}}$, $\rm{Mo_{0.5}Ta_{0.5}}$, $\rm{Nb_{0.25}Ta_{0.75}}$, $\rm{Nb_{0.5}Ta_{0.5}}$, $\rm{Nb_{0.75}Ta_{0.25}}$, $\rm{W_{0.25}Ta_{0.75}}$ and $\rm{W_{0.5}Ta_{0.5}}$.

\begin{figure*}[htbp]
\centering
\includegraphics[width=1.0\linewidth]{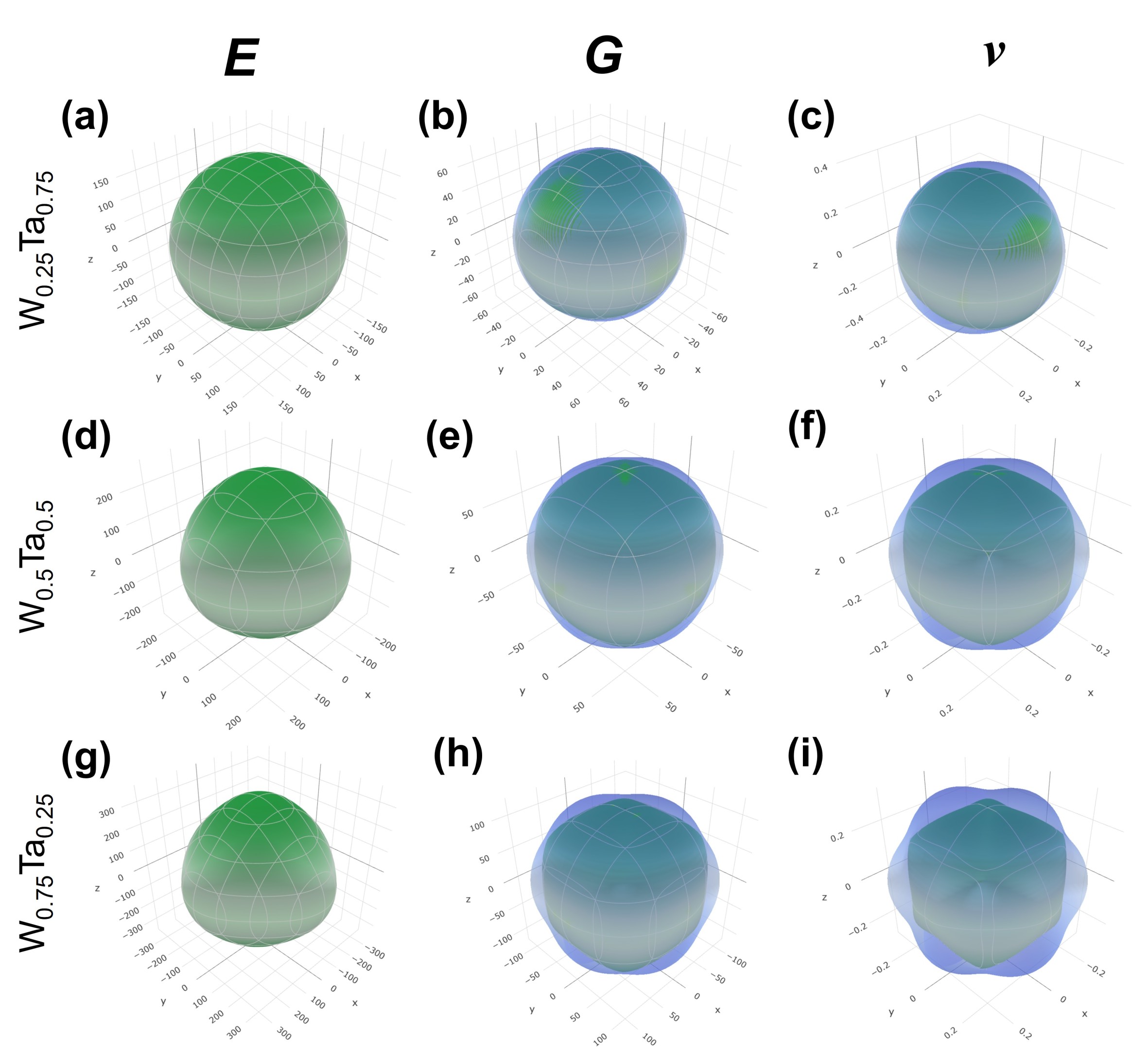}
\caption{\label{fig:anisotropy_WTa} Anisotropy plots of $G$ (in GPa), $E$ (in GPa), and $\nu$ for: (a)-(c)  $\rm{W_{0.25}Ta_{0.75}}$;\, (d)-(f)  $\rm{W_{0.5}Ta_{0.5}}$;\, and (g)-(h) $\rm{W_{0.75}Ta_{0.25}}$. Plots generated using the ELATE tool \cite{gaillac2016elate}. The maximum and minimum values of $G$, $E$ and $\nu$ are represented as transparent blue and solid green surfaces respectively.}
\end{figure*}

\subsection{Nuclear-relaxation contributions to elastic properties}
\label{relaxation-fields}
The method employed in computing the nuclear-relaxation contributions to the elastic tensor makes it possible to decompose it into atom-specific contributions, allowing more detailed modeling of the microscopic origins of elastic response in the binary alloys.Similar to elastic anisotropy plots, the elastic property heterogeneity can be plotted as lattice dependent spatial variations of the elastic properties. The force response internal strain tensor $\Lambda_{p\alpha, j}$ characterizes the force response to the atomic displacement of atom $p$ along direction $\alpha$ when the crystal is under strain ($s_j$), while the displacement response internal strain tensor $\Gamma_{p\alpha, i}$ characterizes the displacement response to the atomic displacement of the atom $p$ along direction $\alpha$ when the crystal is under  strain ($s_i$). The product of these intermediate values summed over all directions for a given atom $p$ gives the relaxation value of a given elastic constant or elastic property ($B$, $G$, $E$ and $\nu$) of atom $p$ under macroscopic strain (Eqs.~(\ref{eq:decomposition}) and (\ref{eq:shear_relaxed})). To illustrate the resulting spatial variation in a relaxed crystal structure, the relaxation effects are displayed as contour plots in Figs.~\ref{fig:forces_NbTa_1}-\ref{fig:forces_WTa_2} for NbTa and WTa, and for MoNb, MoTa, MoW, and WTa in the Appendix. The supercell is represented as a set of $\{ 0 \, 0 \, 1 \}$ planes. A supercell with unit cells stacked in $2\times2\times2$ fashion has 4 unique $\{ 0 \, 0 \, 1 \}$ planes. These are used to construct 2D layers in order to represent the 3D periodic supercell. Each layer contains the atoms of their respective $( 0 \, 0 \, 1 )$ plane, and their first nearest neighbors (1NN) from the next plane. For example, layer 1 consists of atoms from the plane 1 and atoms from plane 2 (these are the 1NN of the atoms in plane 1) projected onto plane 1. Interpolation between atomic contributions is used to construct the spatially-dependent relaxation field \cite{ye2018atomic} for each elastic tensor quantity, using a linear function. The color bar in the contour plot has a range varying from minimum (blue) to maximum (yellow) relaxation value. 

In the case of NbTa (Fig.~\ref{fig:forces_NbTa_1} \& Fig.~\ref{fig:forces_NbTa_2}), the constituent elements Nb and Ta have similar single-bond covalent atomic radii ($r_{\rm cov}$ = 1.47 \text{\AA} for Nb and 1.46 \text{\AA} for Ta \cite{pyykko2009molecular}), and exhibit a small difference in their elastic properties. The relaxation values of the Nb atoms and Ta atoms vary with lattice position for $\bar{C}_{11}$, $\bar{C}_{12}$, $\bar{C}_{44}$, $B$, $G$, $E$ and $\nu$. This is due to differences in the surrounding atoms at each lattice site, which is the nature of disordered systems. The average relaxation value of a Nb atom for $\bar{C}_{11}$, $\bar{C}_{12}$, $\bar{C}_{44}$, $B$, $G$ and $E$ in case of $\rm{Nb_{0.25}Ta_{0.75}}$, where Nb atom is the alloying element, is lower than that of the host Ta atoms while, an inverse behaviour is observed in case of $\nu$. Similarly, the average relaxation value of the Ta atom for $\bar{C}_{11}$, $\bar{C}_{12}$, $\bar{C}_{44}$, $B$, $G$ and $E$ for $\rm{Nb_{0.75}Ta_{0.25}}$, where the Ta atom is alloying element, is lower than that of the host Nb atoms. The distribution of atomic relaxation values for NbTa are similar for $\bar{C}_{11}$, $\bar{C}_{12}$, $\bar{C}_{44}$, $B$, $G$ and $E$ with varying magnitudes except for $\nu$. For $\nu$ the behaviour is observed to be opposite to other elastic properties.

Fig.~\ref{fig:forces_WTa_1} \& ~\ref{fig:forces_WTa_2} shows the decompositions of relaxation values in WTa, for which the constituent elements W and Ta have different single-bond covalent atomic radii ($r_{\rm cov}$ = 1.37 \text{\AA} for W and 1.46 \text{\AA} for Ta \cite{pyykko2009molecular}), and significant differences in their elastic constants. As for WTa, the relaxation values of the W and Ta atoms vary with lattice position for $\bar{C}_{11}$, $\bar{C}_{12}$, $\bar{C}_{44}$, $B$, $G$, $E$ and $\nu$. Similar to NbTa, the average relaxation value of the W atoms is lower for $\bar{C}_{11}$, $\bar{C}_{44}$, $B$, $G$, $E$ and $\nu$ in the case of $\rm{W_{0.25}Ta_{0.75}}$ when W is the alloying element, and the average relaxation value of the Ta atoms is lower for $\bar{C}_{11}$, $\bar{C}_{44}$, $B$, $G$, $E$ and $\nu$ in case of $\rm{W_{0.75}Ta_{0.25}}$ when Ta is the alloying element. However, for $\bar{C}_{12}$ of WTa, the average relaxation value of the W atoms is greater when W is the alloying element ($\rm{W_{0.25}Ta_{0.75}}$) and the average relaxation value of the Ta atoms is greater in $\rm{W_{0.75}Ta_{0.25}}$, where Ta is alloying element.

\begin{figure*}[htbp]
\centering
\includegraphics[width=1.0\linewidth]{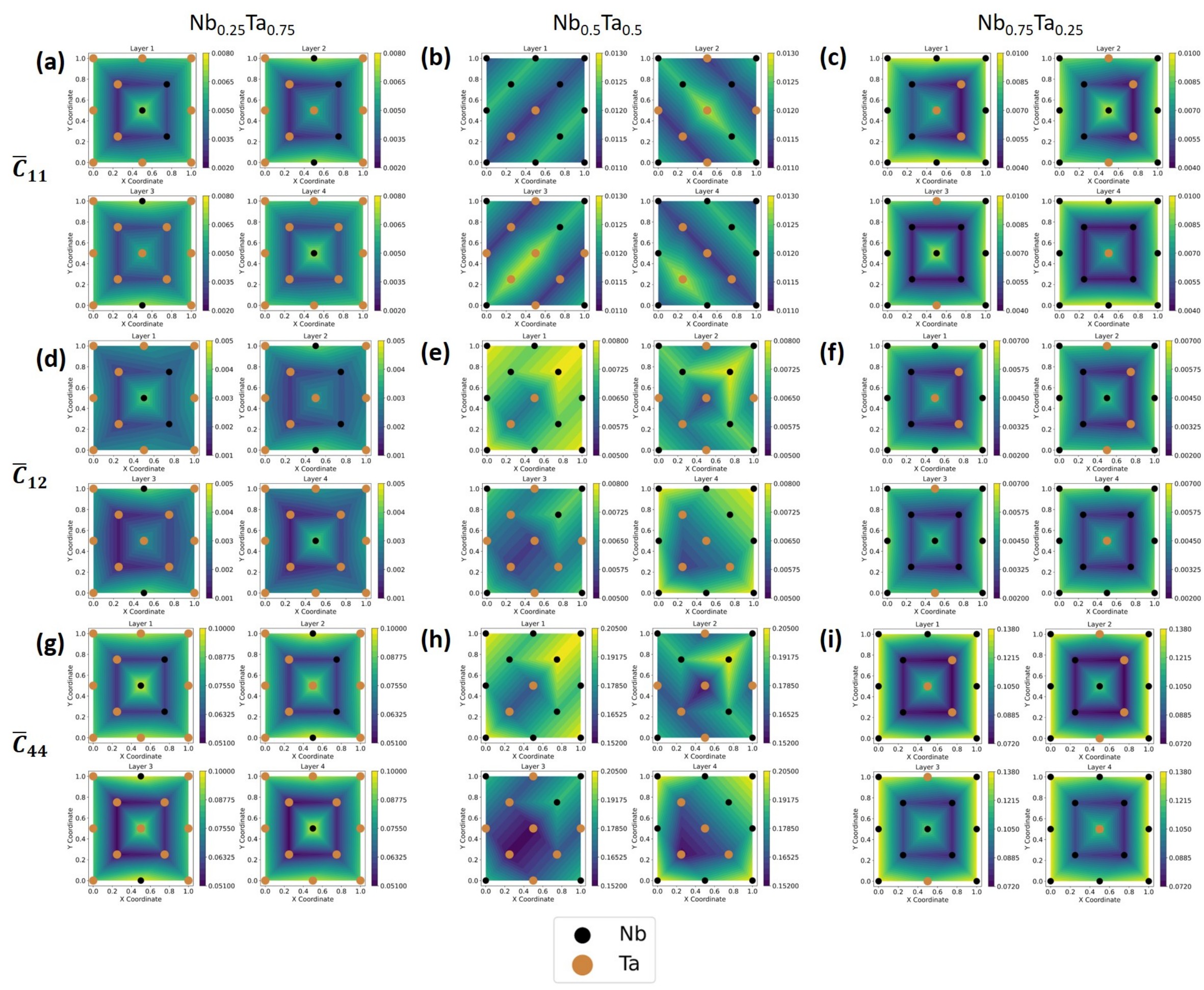}
\caption{\label{fig:forces_NbTa_1} Relaxation field (values in GPa) constructed as described in text for $\rm{Nb_{0.25}Ta_{0.75}}$, $\rm{Nb_{0.5}Ta_{0.5}}$, and $\rm{Nb_{0.75}Ta_{0.25}}$. (a)-(c): $\bar{C}_{11}$,\, (d)-(f): $\bar{C}_{12}$ and (g)-(i): $\bar{C}_{44}$. The color bar is set to range from minimum to maximum relaxation values for all atoms. $x$ and $y$ coordinates correspond to reduced coordinates in the 3D supercell.}
\end{figure*}

\begin{figure*}[htbp]
\centering
\includegraphics[width=1.0\linewidth]{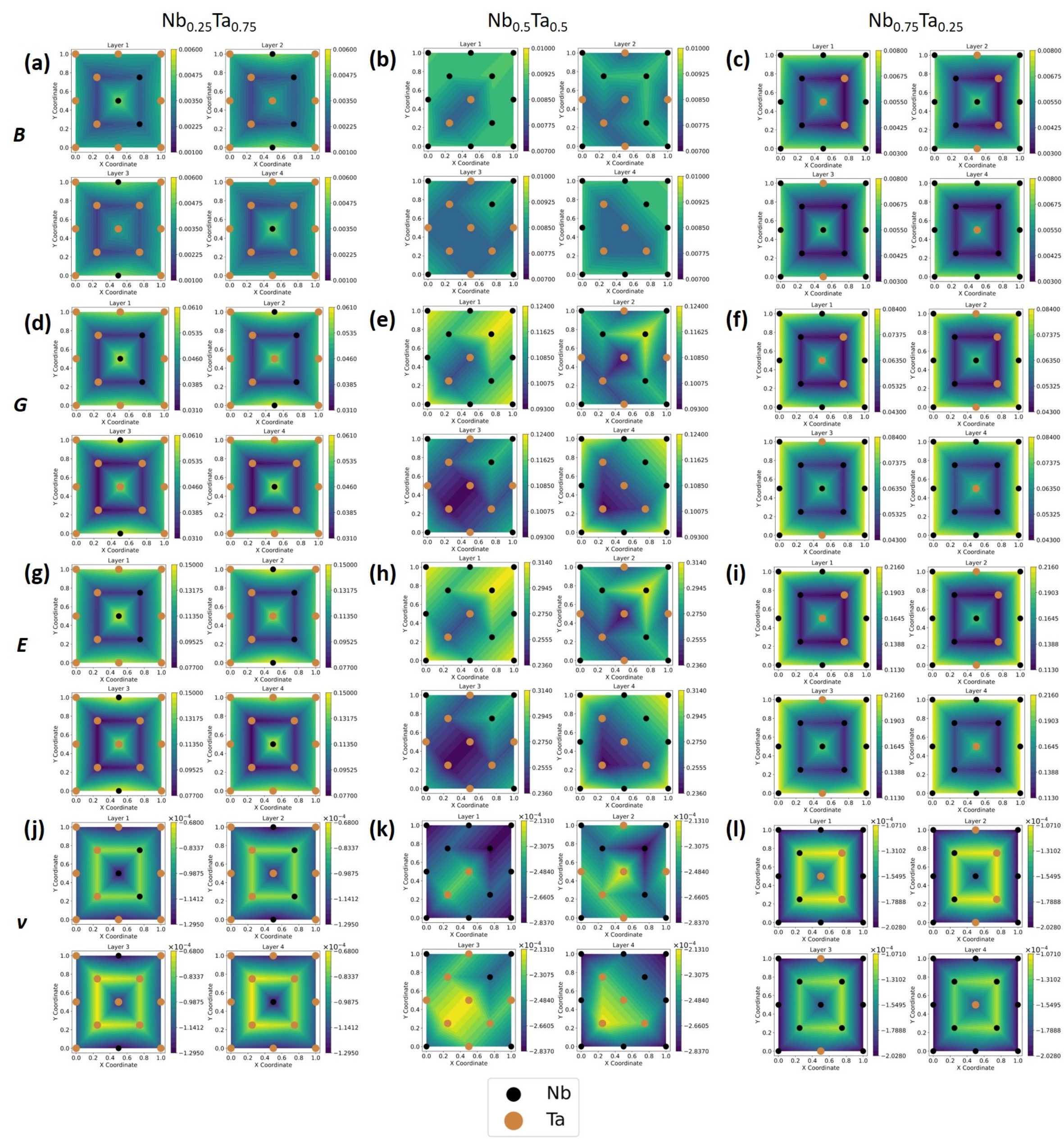}
\caption{\label{fig:forces_NbTa_2} Relaxation field (values in GPa) constructed as described in text for $\rm{Nb_{0.25}Ta_{0.75}}$, $\rm{Nb_{0.5}Ta_{0.5}}$, and $\rm{Nb_{0.75}Ta_{0.25}}$. (a)-(c): $B$,\, (d)-(f): $G$,\, (g)-(i): $E$ and (j)-(l): $\nu$. The color bar is set to range from minimum to maximum relaxation values for all atoms. $x$ and $y$ coordinates correspond to reduced coordinates in the 3D supercell.}
\end{figure*}

\begin{figure*}[htbp]
\centering
\includegraphics[width=1.0\linewidth]{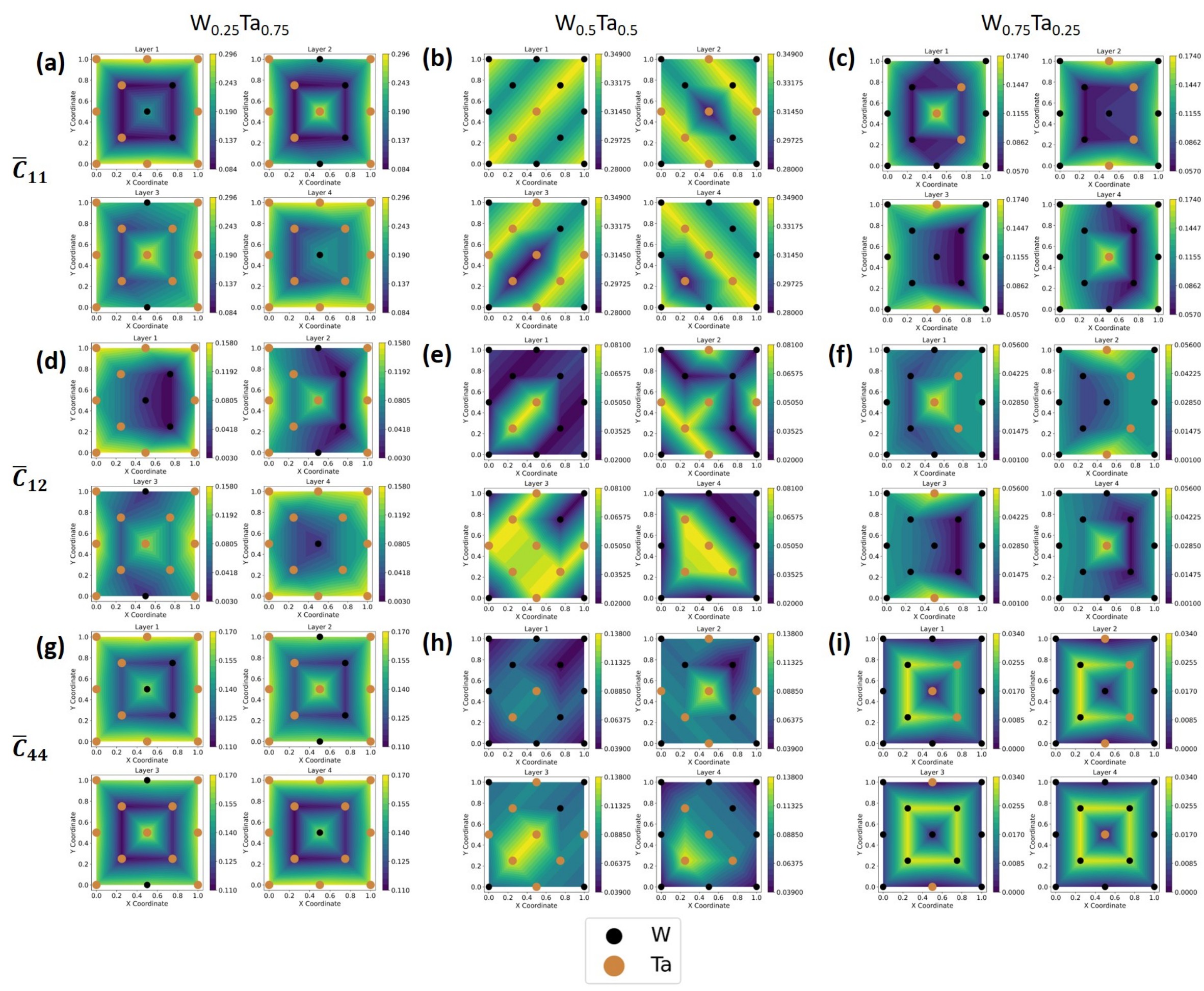}
\caption{\label{fig:forces_WTa_1} Relaxation field (values in GPa) constructed as described in text for $\rm{W_{0.25}Ta_{0.75}}$, $\rm{W_{0.5}Ta_{0.5}}$, and $\rm{W_{0.75}Ta_{0.25}}$. (a)-(c): $\bar{C}_{11}$,\, (d)-(f): $\bar{C}_{12}$ and (g)-(i): $\bar{C}_{44}$. The color bar is set to range from minimum to maximum relaxation values for all atoms. $x$ and $y$ coordinates correspond to reduced coordinates in the 3D supercell.}
\end{figure*}

\begin{figure*}[htbp]
\centering
\includegraphics[width=1.0\linewidth]{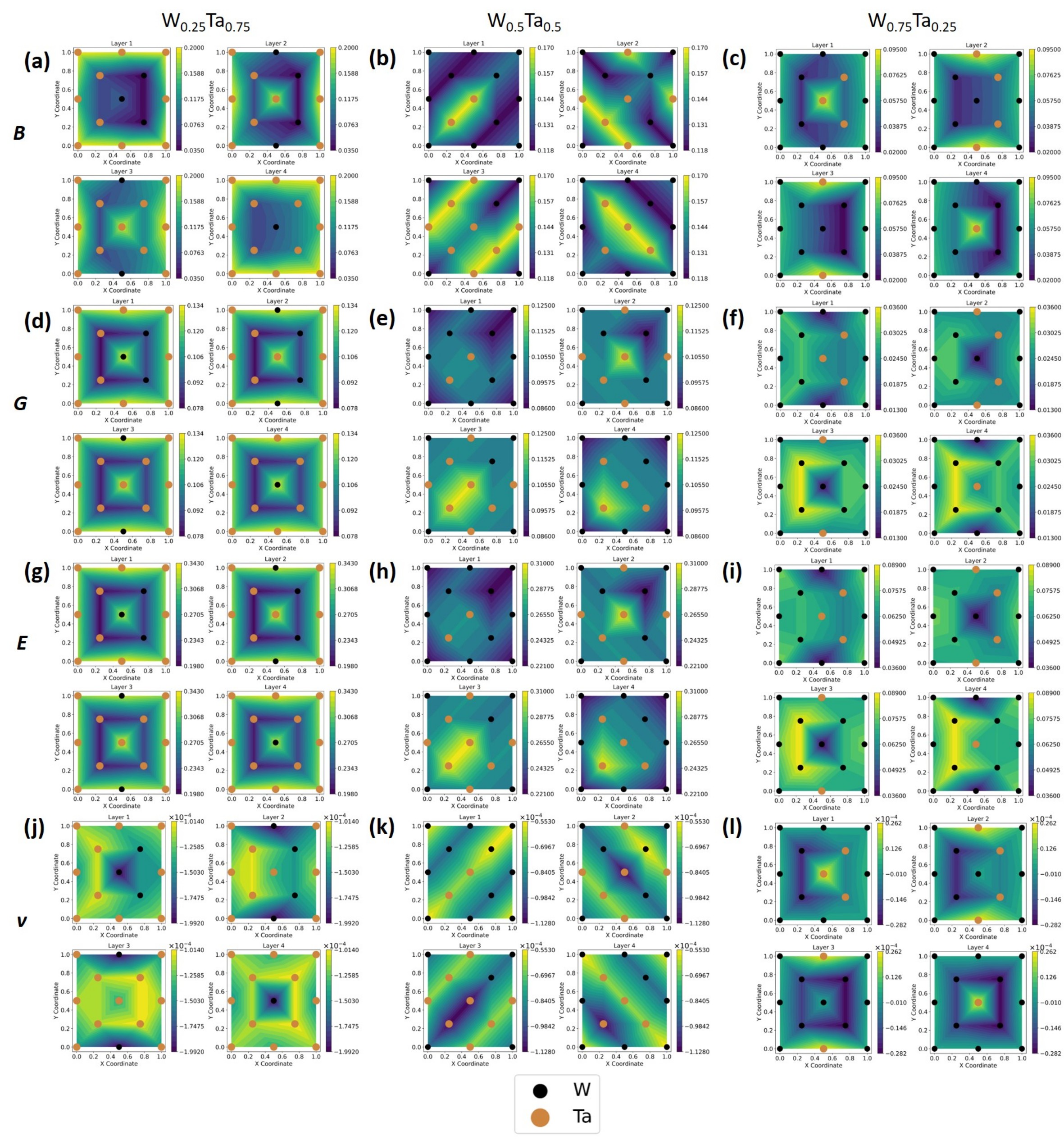}
\caption{\label{fig:forces_WTa_2} Relaxation field (values in GPa) constructed as described in text for $\rm{W_{0.25}Ta_{0.75}}$, $\rm{W_{0.5}Ta_{0.5}}$, and $\rm{W_{0.75}Ta_{0.25}}$. (a)-(c): $B$,\, (d)-(f): $G$,\, (g)-(i): $E$ and (j)-(l): $\nu$. The color bar is set to range from minimum to maximum relaxation values for all atoms. $x$ and $y$ coordinates correspond to reduced coordinates in the 3D supercell.}
\end{figure*}

\section{Discussion}
\label{sec:Discussion}
\subsection{Elastic tensor variations with distortions in lattice}
Within continuum elasticity theory \cite{zhao2007first}, the ideal elastic tensor of a cubic crystal (including BCC) has only three independent elastic constants due to the lattice symmetry of the crystal. The number of independent elastic constants will increase when the symmetry of the crystal is broken. Alloying introduces distortions in the perfect lattice due to differences in the atomic radii and quantum mechanical interactions between the constituent atoms \cite{bijjala2025composition}. This causes lattice distortions and introduces local strain, which hinders dislocation motion, thus increasing the mechanical strength \cite{senkov2010refractory, zhang2014microstructures}. Hence it is crucial to correlate the effects of distortions on the elastic tensor due to alloying, and develop computational methods for characterizing these effects at the microscopic level.

To understand the effect of distortions on the elastic constants we define two types of distortions, crystal distortions and lattice distortions. Crystal distortions correspond to changes in the lattice parameter in different directions due to symmetry breaking of the BCC crystal. Lattice distortions refer to atomic motion away from ideal lattice sites due to differences in the electronic (quantum mechanical) interactions between the atoms. Computation of the relaxed-ion elastic tensor using DFPT, instead of computing only $C_{11}$, $C_{12}$ and $C_{44}$, makes it possible to capture crystal distortions and lattice distortions in terms of standard deviations in Voigt averages and nuclear relaxation values in the BCC binary alloys. Non-zero standard deviation values in the Voigt-averaged elastic constants signal crystal distortions ${\rm a}_x \neq {\rm a}_y \neq {\rm a}_z$. In the case of A = 50 at.\% composition in the AB binary alloys, the relaxed structures retain cubic crystal symmetry (${\rm a}_x ={\rm a}_y={\rm a}_z$), leading to ${C}_{11} = {C}_{22}={C}_{33}$, ${C}_{12} = {C}_{23}={C}_{31}$ and ${C}_{44} = {C}_{55}={C}_{66}$, and thus a zero standard deviation in the Voigt-averaged values. Nevertheless, although crystal symmetry is preserved in the 50-50 binary alloys, deviations away from the perfect lattice are still possible, and indeed, these can play an important role in determining the mechanical behavior of high-entropy alloys \cite{owen2018lattice}.

\begin{figure*}[htbp]
\centering
\includegraphics[width=0.725\linewidth]{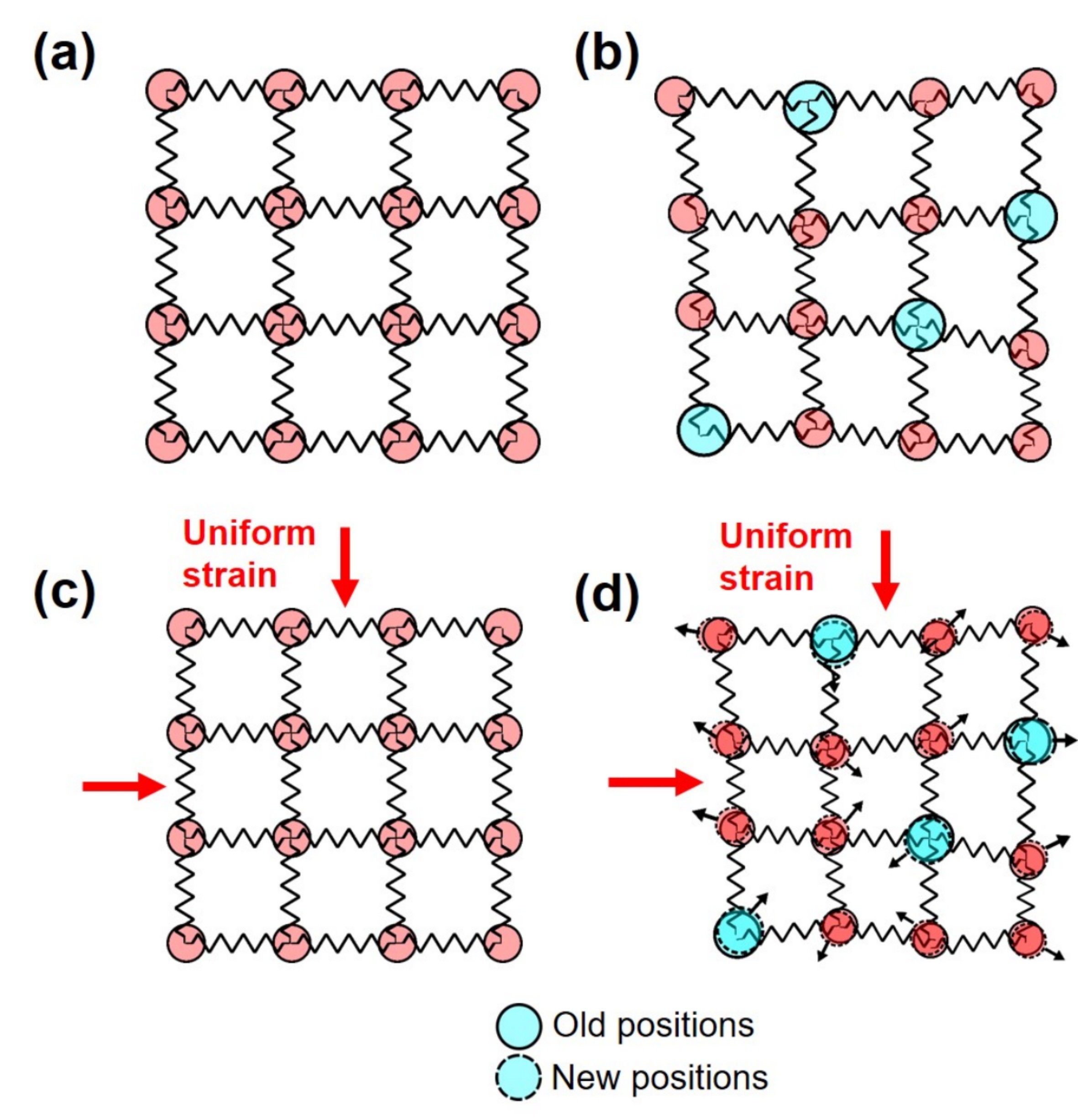}
\caption{\label{fig:overall} Resultant forces on atoms under uniform strain in unary and binary alloys. ($\boldsymbol{a}$) Bulk crystal in relaxed state; ($\boldsymbol{b}$) binary alloy in relaxed state; ($\boldsymbol{c}$) bulk crystal in strain state; and ($\boldsymbol{d}$) binary alloy in strain state. Here dashed circles represent the new position of atoms in a strained crystal in its relaxed state. Red arrows indicates the deformation direction; black arrows indicate resultant forces on the atoms.}
\end{figure*}

Lattice distortions in the relaxed structure of the binary alloys lead to unbalanced forces on the crystal under a finite strain as illustrated schematically in  Fig.~\ref{fig:overall}. The magnitude and direction of resultant unbalanced forces at each lattice site depends on the atom type and the atomic environment in the lattice. In disordered binary alloys, atom type and atomic environment differ at each lattice site. Nuclear relaxation values, which contribute to the computation of relaxed-ion elastic constants, provides the measure of resultant unbalanced forces in the crystal under finite strain. While standard deviation values for the binary alloys with A = 50 at.\% is zero, indicating cubic crystal structure, the nuclear relaxation values are non-zero and are maximum for $\bar{C}_{11}$. This indicates that in A = 50 at.\% binary alloys, even though there are no crystal distortions, there are lattice distortions in the crystal which lead to unbalanced forces in the crystal. For same-group binary alloys, $\bar{C}_{11}$, $\bar{C}_{12}$ and $\bar{C}_{44}$ are observed to be close to zero indicating minute lattice distortions except for $\bar{C}_{44}$ of NbTa. For $\bar{C}_{44}$ of NbTa, the nuclear relaxation values are finite even though the standard deviation values are zero. This indicates that, in NbTa, the unbalanced forces are oriented in a particular direction as only $\bar{C}_{44}$ has a finite contribution, while $\bar{C}_{11}$ and $\bar{C}_{12}$ has near-zero contributions. Even though on the macroscopic scale, no crystal distortions are observed, at the microscopic scale the unbalanced forces are oriented in a particular direction, which reduces the magnitude of relaxed-ion elastic constants. In case of MoW and NbTa, nuclear relaxation values are near zero (except for $\bar{C}_{44}$ of NbTa) for all compositions, while the standard deviation values are finite for A = 25 at.\% and 50 at.\% compositions. This may be due to near-zero unbalanced forces on the atoms under strain or random orientation of the forces on the atoms resulting in near-zero overall relaxation values. 

\subsection{Effect of alloying on elastic properties}
The input structures used in the DFPT rigid-ion and relaxed-ion calculations can impact the accuracy of the final computed elastic tensors, through residual stresses. In principle, it is possible to correct for any initial stresses present in the starting structure \cite{romero2020abinit}. In the present work, we have utilized fully-relaxed input SQS supercell structures in order to avoid the need for {\it ex post facto} corrections and to maintain internal consistency in our calculations. The input structures are obtained for each binary and composition by minimizing the forces on each atom, allowing cell shape, dimensions and relative atomic positions on the lattice to vary.  The particular computational procedures followed here have been essential to obtaining elastic constants in remarkably good agreement with experiment. 

Elastic constants are fundamental probe of the elastic interactions of atoms in a given material and are correlated with mechanical properties. Due to random arrangement of atoms in disordered alloys, the elastic interactions are complex and difficult to quantify. In single crystal disordered alloys, solid solution effects like solid solution strengthening are the major contributing factors to observed changes in mechanical properties due to alloying. Since the seminal works of Fleisher \cite{fleischer1963substitutional} and Labusch \cite{labusch1970statistical}, solid solution effects have been correlated with differences in atomic radii and elastic moduli. The fundamental idea behind this is to capture the lattice distortions and associated local strains due to alloying \cite{he2018lattice,toda2015modelling}. Even though these models have been widely verified for yield strength, they can be used to understand trends in elastic constants as well \cite{tian2015elastic, delczeg2009ab}. For same-group binary alloys (MoW and NbTa), the differences in single-bond covalent atomic radii and differences in shear moduli of the constituent elements are smaller compared to those of different-group elements. Hence same-group alloys showed a linear trend across the entire composition range for $\bar{C}_{11}$, $\bar{C}_{12}$ and $\bar{C}_{44}$, closely following Vegard's law. However, in different-group binary alloys (MoNb, MoTa, WNb and WTa) the differences in atomic radii and differences in shear moduli of the constituent elements lead to nonlinear trends, with deviations from Vegard's law.

\subsection{Nuclear relaxation contributions to the elastic tensor }
Nuclear-relaxation contributions to the elastic constants capture the influence of atomic rearrangement caused by alloying. These can be physically interpreted as a measure of resultant unbalanced forces on the atoms due to lattice distortions when the crystal is under strain. Relaxation values corresponding to different elastic constants describe relaxation effects in a particular direction of the atoms under strain. In disordered random alloys, the chemical randomness at each lattice site leads to differences in relaxation values at each lattice site. This value is subtracted from the rigid-ion elastic constant values, which are uniform in a crystal, in order to obtain the physical elastic constants. Hence relaxation values provide a measure of heterogeneity in elastic constants.

The elastic constant trends showed deviations from Vegard's law in case of different-group binary alloys. A positive deviation (value greater than value computed using Vegard's law) is seen in the compositions with higher atomic fraction of the constituent element with higher elastic constant value in the case of $\bar{C}_{11}$ and $\bar{C}_{44}$, while an inverse trend is found for $\bar{C}_{12}$. In case of different-group binary alloys, average relaxation values for alloying element is lower than the average relaxation values of the host element for $\bar{C}_{11}$ and $\bar{C}_{44}$. For $\bar{C}_{12}$ the trend is exactly opposite, similar to the trends in deviations from Vegard's law.

Elastic constants are fundamental probes of the nature of bonding in a given material. Elastic constant heterogeneity provides a direct measure of varying bond strength in alloys. Higher relaxation values lead to lower elastic constant values, as relaxation values are subtracted from the rigid-ion elastic constants. This effect plays a crucial role in understanding the location of a crack's initial site in disordered alloys, as locations with higher relaxation values will have lower elastic constants, hence higher stress concentration sites, thus serving as crack initiation sites. The methodology demonstrated here can be easily extended to multi-component systems where the relaxation values are expected to be even higher due to larger lattice distortions \cite{miracle2017critical}.  

\section{Conclusions}
\label{sec:Conclusions} 
In this work, the elastic tensors of composition-dependent refractory binary alloys based on Mo, Nb, Ta, and W have been computed 
using density functional perturbation theory in conjunction with SQS supercell modeling of the composition-dependent disordered BCC structures. The use of the response-based DFPT methodology for computing the elastic tensor made it possible to compute both the rigid-ion and relaxed ion elastic tensors and nuclear-relaxation contributions, in addition to elastic tensor-derived properties. The Voigt-averaged elastic constants $\bar{C}_{11}, \bar{C}_{12}$, and $\bar{C}_{44}$ derived from the elastic tensors were used to compute the polycrystalline elastic moduli $E$, $B$, and $G$.  Relaxed-ion calculations for the elastic constants corresponding to the inclusion of nuclear relaxation effects under uniform strain are consistent with experiment. Differences between the rigid-ion and relaxed-ion values for the elastic constants were found to lie within the standard deviations of their Voigt averages for the 25, 50, and 75 at.\% alloy compositions, suggesting that nuclear relaxation contributions to the elastic tensor are relatively small.  This made it possible to extend rigid-ion elastic tensor and derived elastic properties to additional alloy compositions with no loss in accuracy. The computed rigid-ion elastic constants are in excellent agreement with available experimental data, demonstrating the ability of our methodology to accurately predict key mechanical properties from first-principles theory. Derived mechanical properties of $B$, $G$, and $E$ showed a linear trend across the compositions for same-group binary alloys and a nonlinear trend in different-group binary alloys. Our calculations showed elastic anisotropy behavior in all of the binary alloys except WTa.

The computed relaxation values were shown to be small and the decomposition of the elastic constants into electronic contributions and nuclear relaxation contributions helps to understand the effects of nuclear relaxations on the symmetry of the elastic tensor. The lattice-dependent nuclear relaxation fields provide detailed quantum mechanical information for potential use in data-driven predictive models of more complex alloys (ternary and beyond).  Such models will be essential in order to overcome computational bottlenecks and scaling issues in property-driven materials design as system sizes grow and compositional complexity increases. Computation of relaxation values made it possible to map the heterogeneity in elastic constants which can play a crucial role in understanding crack initial site in disordered alloys. This demonstrated ability to perform accurate, experimentally-validated DFPT calculations will enable advanced, microscopic modeling of the mechanical properties of compositionally- and chemically-complex alloys. 

\section*{CRediT authorship contribution statement}
{\bf Surya Bijjala:} Methodology, Software, Visualization, Formal analysis, Data curation, Writing – original draft. {\bf Susan R.~Atlas:} Conceptualization, Methodology, Formal analysis, Supervision, Writing – review \& editing. {\bf Pankaj Kumar:} Conceptualization, Formal analysis, Supervision, Writing - Review \& Editing, Funding acquisition. 

\section*{Declaration of competing interest}
The authors declare that they have no known competing financial interests or personal relationships that could have appeared to influence the work reported in this paper.

\section*{Acknowledgements}
This work was supported in part by NASA EPSCoR Grant \# 80NSSC21M0171. We would like to thank the UNM Center for Advanced Research Computing, supported in part by the National Science Foundation, for providing computational resources and system support for this work. Comments and suggestions by anonymous reviewers are gratefully acknowledged. 

\section*{Data Availability}
The data that support the findings of this study are available from the corresponding authors upon reasonable request.

\bibliography{RHEAMechanicalProp}

\appendix
\section*{Appendix}
\setcounter{section}{1}
\setcounter{equation}{0}
\renewcommand{\theequation}{A.\arabic{equation}}

\vspace*{.05in}
\subsection{Derivation of nuclear relaxation expressions for Young's modulus $E$ and Poisson ratio $\nu$.}

\vspace*{.05in}
\noindent

{\bf Young's modulus $E$.} Starting from Eq.~(\ref{eq:Young's_relaxed}), and the Voigt expression for $E$ (Eq.~(\ref{eq:E_voigt})), we want to derive an expression for $E_\text{nuclear}$ in terms of the quantities $\{B_n,G_n,B_r,G_r\}$, where $B_n = B_{\text{nuclear}}$, $G_n = G_{\text{nuclear}}$, $B_r = B_{\text{rigid}}$, and $G_r = G_{\text{rigid}}$ as defined in the main text.

\vspace*{.1in}
\noindent
We have:
\begin{align} 
\label{eq:E_voigt_derivation}
E_{\text{relaxed}} = E_{\text{Voigt}} &= \frac{9B_{\text{Voigt}}G_{\text{Voigt}}}{3B_{\text{Voigt}}+G_{\text{Voigt}}} 
= \frac{9 \left( B_r - B_n \right)\left( G_r - G_n \right)}{3\left( B_r - B_n \right)+\left( G_r - G_n \right)} \nonumber \\
&= \frac{9 \left( B_rG_r-B_rG_n-B_nG_r+B_nG_n \right)}{\left( 3B_r + G_r \right)-\left( 3B_n + G_n \right)} 
= \frac{9B_rG_r - 9\left(B_r G_n + B_n G_r - B_n G_n \right)}{\left( 3B_r + G_r \right)\left( 1 - E_1\right)}, \nonumber
\end{align}
where we have defined
\begin{equation}
\label{eq:E_1}
E_1 \equiv \frac{3B_n + G_n}{3B_r + G_r}.
\end{equation}
If the nuclear relaxation contributions to $B$ and $G$ are assumed to be small compared to the rigid values for $B$ and $G$, then $E_1 \ll 1$; this condition will be shown to be an excellent approximation in Section A.2 below.  We can therefore approximate $1/(1-E_1)= (1-E_1)^{-1}$ by its first order Taylor series value, $1+E_1$. Noting that 
\[
E_{\text{rigid}} = \frac{9B_rG_r}{3B_r + G_r},
\]
and defining
\begin{equation}
E_2 \equiv \frac{B_r G_n +B_n G_r - B_n G_n}{B_r G_r},
\end{equation}
we have:
\begin{align}
E_{\text{Voigt}} &= \frac{9B_rG_r - 9\left(B_r G_n + B_n G_r - B_n G_n \right)}{\left( 3B_r + G_r \right)\left( 1 - E_1\right)}
\nonumber \\
&\approx \left (\frac{9B_rG_r}{3B_r + G_r} - \frac{9\left(B_r G_n + B_n G_r - B_n G_n \right)}{3B_r + G_r} \right ) \left (1+E_1 \right ) \nonumber \\
&= \frac{9B_rG_r}{3B_r + G_r} \left ( 1 - \frac{\left (B_r G_n + B_n G_r - B_n G_n\right)}{B_rG_r}\right ) \left (1+E_1 \right ) \nonumber \\
&= E_{\text{rigid}}(1-E_2)(1+E_1) = E_{\text{rigid}} (1-E_2+E_1-E_2E_1) \nonumber \\
&= E_{\text{rigid}} - E_{\text{rigid}}(E_2 - E_1 + E_2E_1). \nonumber
\end{align}
Thus,
\begin{equation}
    E_{\text{nuclear}} \approx E_{\text{rigid}}(E_2 - E_1 + E_2E_1).
\end{equation}

\vspace*{.25in}
\noindent
{\bf Possion's ratio $\nu$}.  Starting from Eq.~(\ref{eq:Possion's_relaxed}):
\begin{align}
\nu_{\text{relaxed}} = \nu_\text{Voigt} &= \frac{3B_{\text{Voigt}}-2G_{\text{Voigt}}}{2\left(3B_{\text{Voigt}}+G_{\text{Voigt}}\right)} 
= \frac{3\left(B_r - B_n \right)-2\left(G_r - G_n \right)}{2\left(3\left(B_r - B_n \right)+\left(G_r - G_n \right)\right)} \nonumber \\
&= \frac{\left(3B_r-2G_r\right) - \left( 3B_n - 2G_n\right)}{2\left(3B_r+G_r\right)-2\left(3B_n+G_n\right)} 
= \frac{\left(3B_r-2G_r\right) - \left( 3B_n - 2G_n\right)}{2\left(3B_r+G_r\right) \left(1- \nu_1 \right)}, \nonumber
\end{align}
where we have defined
\begin{equation}
\label{eq:v_1}
    \nu_1 =
    \frac{3B_n + G_n}{3B_r + G_r}.
\end{equation}
Note that $\nu_1 = E_1$ from the Young's modulus derivation.  We again assume that $\nu_1 \ll 1$, so that $1/(1-\nu_1) \approx 1+ \nu_1$. This gives:
\begin{align}
\nu_\text{Voigt} &\approx \left ( \frac{\left(3B_r-2G_r\right) - \left( 3B_n - 2G_n\right)}{2\left(3B_r+G_r\right) } \right) (1+\nu_1) = \left(\frac{3B_r-2G_r}{2(3B_r + G_r)} - \frac{(3B_n - 2G_n)}{2(3B_r+G_r)} \right ) (1+\nu_1) \nonumber \\
&= \nu_{\text{rigid}} \left (1-\nu_2 \right) (1+\nu_1) = \nu_{\text{rigid}} \left (1-\nu_2 + \nu_1 - \nu_2\nu_1\right) \nonumber \\
&= \nu_{\text{rigid}} - \nu_{\text{rigid}} \left (\nu_2 - \nu_1 + \nu_2\nu_1\right ), \nonumber
\end{align}
where we have defined

\begin{equation}
    \nu_2 = \frac{3B_n - 2G_n}{3B_r - 2G_r}.
\end{equation}
Thus,
\begin{equation}
    \nu_{\text{nuclear}} \approx \nu_{\text{rigid}} \left (\nu_2 - \nu_1 + \nu_2\nu_1\right ).
\end{equation}

\vspace*{.05in}
\subsection{Nuclear relaxation field validation tables}
\vspace*{.05in}
\noindent
\begin{table*}[!ht]
    \centering
    \caption{Comparison of Abinit-computed nuclear-relaxation values ($E_{nuclear}$) with summation of approximated atomic nuclear relaxation values ($\sum_p E_{p,nuclear}$) over all atoms $p$ in the supercell. For $E_1$ the max value corresponds to maximum value of $E_1$ defined in Eq.(~\ref{eq:E_1}) across all atoms in the supercell.}
    \scalebox{0.7}{
    \begin{tabular}{|c|c|c|c|c|c|c|c|c|c|}
        \hline
        \multicolumn{2}{|c|}{\textbf{Composition}} & \multirow{2}{*}{\begin{tabular}{@{}c@{}}$E_1$\\ (max value)\end{tabular}} & \multirow{2}{*}{$E_{nuclear}$} & \multirow{2}{*}{$\sum_p E_{p,nuclear}$} & 
        \multicolumn{2}{|c|}{\textbf{Composition}} & \multirow{2}{*}{\begin{tabular}{@{}c@{}}$E_1$ \\ (max value)\end{tabular}} & \multirow{2}{*}{$E_{nuclear}$} & \multirow{2}{*}{$\sum_p E_{p,nuclear}$} \\
        \cline{1-2} \cline{6-7}
        AB & at.\% of A                            &                        &                        & &
        AB & at.\% of A                            &                        &                       & \\
        \hline
        \multirow{3}{*}{MoNb} &  25  & 8.16E-04&2.7267 &2.7253 & \multirow{3}{*}{MoTa} &  25  & 1.07E-03&9.2285 &9.1913 \\
        &50 & 5.24E-04&1.9929&1.9933&  &50 & 5.19E-04&5.1921&5.1829 \\
        &75 & 3.30E-04&1.0192&1.0193& &75 & 2.48E-04&1.3723&1.3718  \\
        \hline

        \multirow{3}{*}{MoW} &  25  & 6.91E-05&0.2084&0.2084 & \multirow{3}{*}{NbTa} &  25  & 1.22E-04&1.7917&1.7898 \\
        &50 & 1.00E-04&0.3951&0.3951& &50 & 2.57E-04&4.3686&4.3573 \\
        &75 & 6.98E-05&0.1786&0.1786& &75 & 1.86E-04&2.6172&2.6130 \\
        \hline
        \multirow{3}{*}{WNb} &  25  & 9.75E-04&1.1373&1.1384 & \multirow{3}{*}{WTa} &  25  & 9.98E-04&4.3394&4.3367 \\
        &50 & 9.57E-04&2.1130&2.1150& &50 & 7.37E-04&4.1908&4.1907 \\
        &75 & 4.21E-04&1.0581&1.0584& &75 & 3.24E-04&1.0794&1.0796 \\
        \hline
    \end{tabular}}
    \label{tab:nuclear-relxation comparison for E}
\end{table*}

\begin{table*}[!ht]
    \centering
    \caption{Comparison of Abinit-computed nuclear-relaxation values ($\nu_{nuclear}$) with summation of approximated atomic nuclear relaxation values ($\sum_p \nu_{p,nuclear}$) over all atoms $p$ in the supercell. For $\nu_1$ the max value corresponds to maximum value of $\nu_1$ defined in Eq.(~\ref{eq:v_1}) across all atoms in the supercell.}
    \scalebox{0.7}{
    \begin{tabular}{|c|c|c|c|c|c|c|c|c|c|}
        \hline
        \multicolumn{2}{|c|}{\textbf{Composition}} & \multirow{2}{*}{\begin{tabular}{@{}c@{}}$\nu_1$\\ (max value)\end{tabular}} & \multirow{2}{*}{$\nu_{nuclear}$} & \multirow{2}{*}{$\sum_p \nu_{p,nuclear}$} & 
        \multicolumn{2}{|c|}{\textbf{Composition}} & \multirow{2}{*}{\begin{tabular}{@{}c@{}}$\nu_1$ \\ (max value)\end{tabular}} & \multirow{2}{*}{$\nu_{nuclear}$} & \multirow{2}{*}{$\sum_p \nu_{p,nuclear}$} \\
        \cline{1-2} \cline{6-7}
        AB & at.\% of A                            &                        &                        & &
        AB & at.\% of A                            &                        &                       & \\
        \hline
        \multirow{3}{*}{MoNb} &  25  & 8.16E-04&-1.616E-03 &-1.603E-03 & \multirow{3}{*}{MoTa} &  25  & 1.07E-03&-6.716E-03 &-6.652E-03 \\
        &50 & 5.24E-04&-5.552E-04&-5.510E-04&  &50 & 5.19E-04&-3.175E-03&-3.157E-03 \\
        &75 & 3.30E-04&-2.688E-04&-2.680E-04&  &75 & 2.48E-04&-7.135E-04&-7.123E-04  \\
        \hline

        \multirow{3}{*}{MoW} &  25  & 6.91E-05&3.741E-05&3.739E-05 & \multirow{3}{*}{NbTa} &  25  & 1.22E-04&-1.560E-03&-1.558E-03 \\
        &50 & 1.00E-04&7.868E-05&7.861E-05& &50 & 2.57E-04&-3.952E-03&-3.939E-03 \\
        &75 & 6.98E-05&4.592E-05&4.590E-05& &75 & 1.86E-04&-2.465E-03&-2.459E-03 \\
        \hline
        \multirow{3}{*}{WNb} &  25  & 9.57E-04&4.534E-04&4.520E-04 & \multirow{3}{*}{WTa} &  25  & 9.75E-04&4.534E-04&4.520E-04 \\
        &50 & 9.57E-04&8.572E-04&8.501E-04& &50 & 7.37E-04&-1.290E-03&-1.276E-03 \\
        &75 & 4.21E-04&2.917E-04&2.909E-04& &75 & 3.24E-04&-6.232E-05&-6.192E-05 \\
        \hline
    \end{tabular}}
    \label{tab:nuclear-relxation comparison for nu}
\end{table*}

\vspace*{.05in}
\subsection{Binary alloy anisotropy plots and nuclear relaxation fields} 
\begin{figure*}[htbp]
\centering
\includegraphics[width=1.0\linewidth]{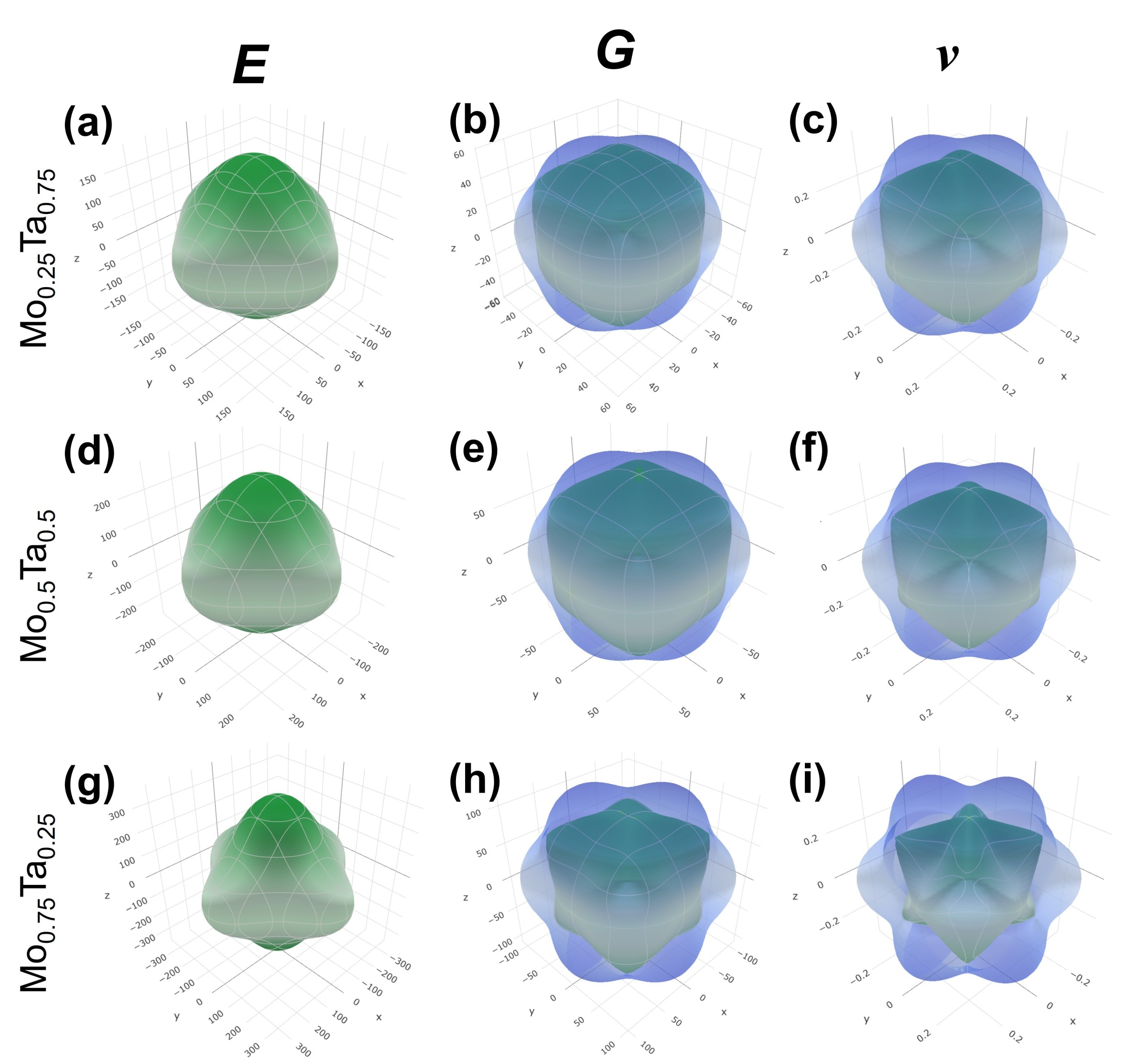}
\caption{\label{fig:anisotropy_MoTa} Anisotropy plots of $G$ (GPa), $E$ (GPa), and $\nu$ for: (a)-(c)\, $\rm{Mo_{0.25}Ta_{0.75}}$;\, (d)-(f) $\rm{Mo_{0.5}Ta_{0.5}}$; and (g)-(h) $\rm{Mo_{0.75}Ta_{0.25}}$. Plots generated using the ELATE tool \cite{gaillac2016elate}. The maximum and minimum values of $G$, $E$ and $\nu$ are represented as transparent blue and solid green surfaces respectively.}
\end{figure*}

\begin{figure*}[htbp]
\centering
\includegraphics[width=1.0\linewidth]{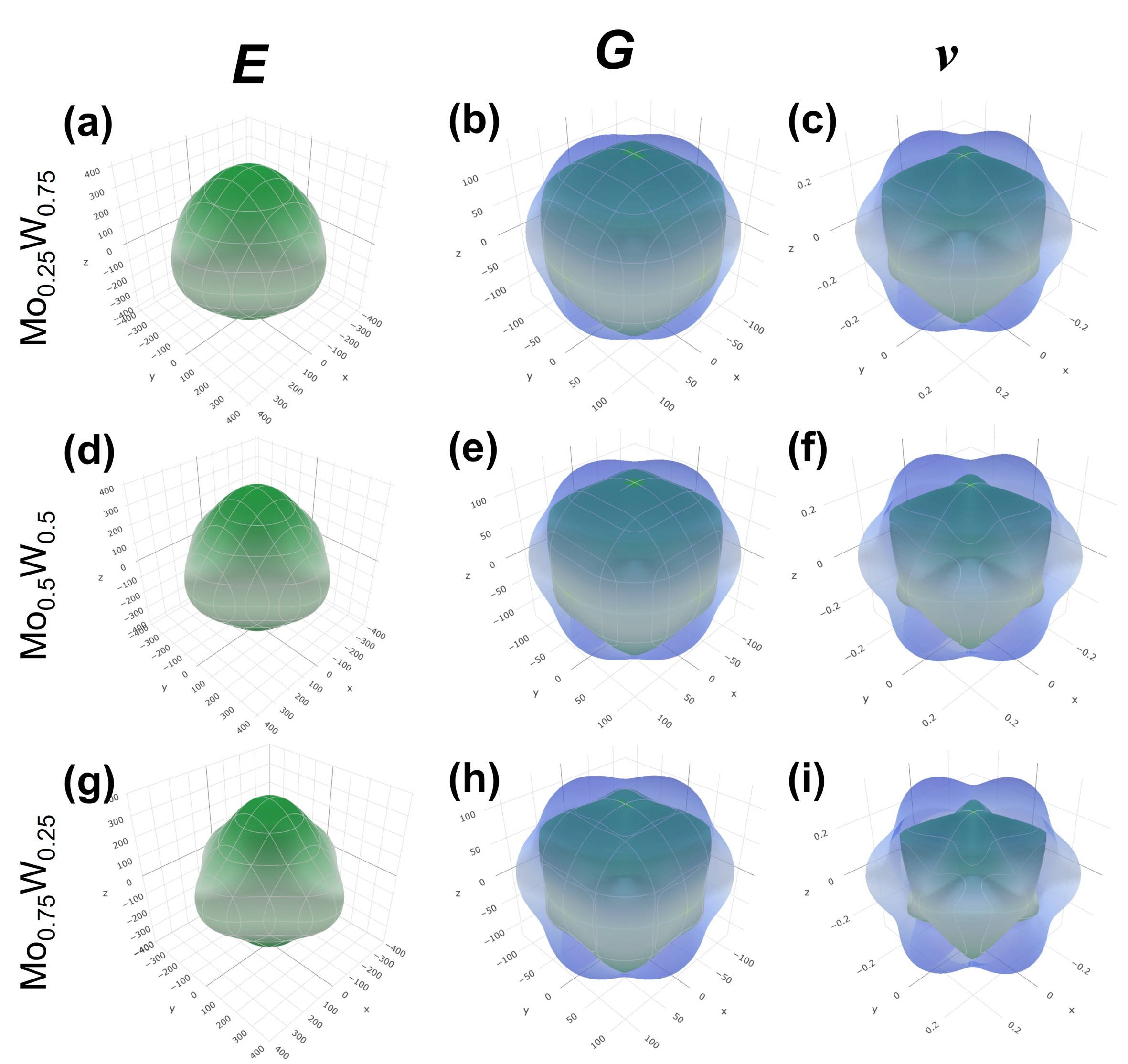}
\caption{\label{fig:anisotropy_MoW} Anisotropy plots of $G$ (GPa), $E$ (GPa) and $\nu$ for: (a)-(c) $\rm{Mo_{0.25}W_{0.75}}$;\, (d)-(f) $\rm{Mo_{0.5}W_{0.5}}$; and (g)-(h) $\rm{Mo_{0.75}W_{0.25}}$. Plots generated using the ELATE tool \cite{gaillac2016elate}. The maximum and minimum values of $G$, $E$ and $\nu$ are represented as transparent blue and solid green surfaces respectively.}
\end{figure*}

\begin{figure*}[htbp]
\centering
\includegraphics[width=1.0\linewidth]{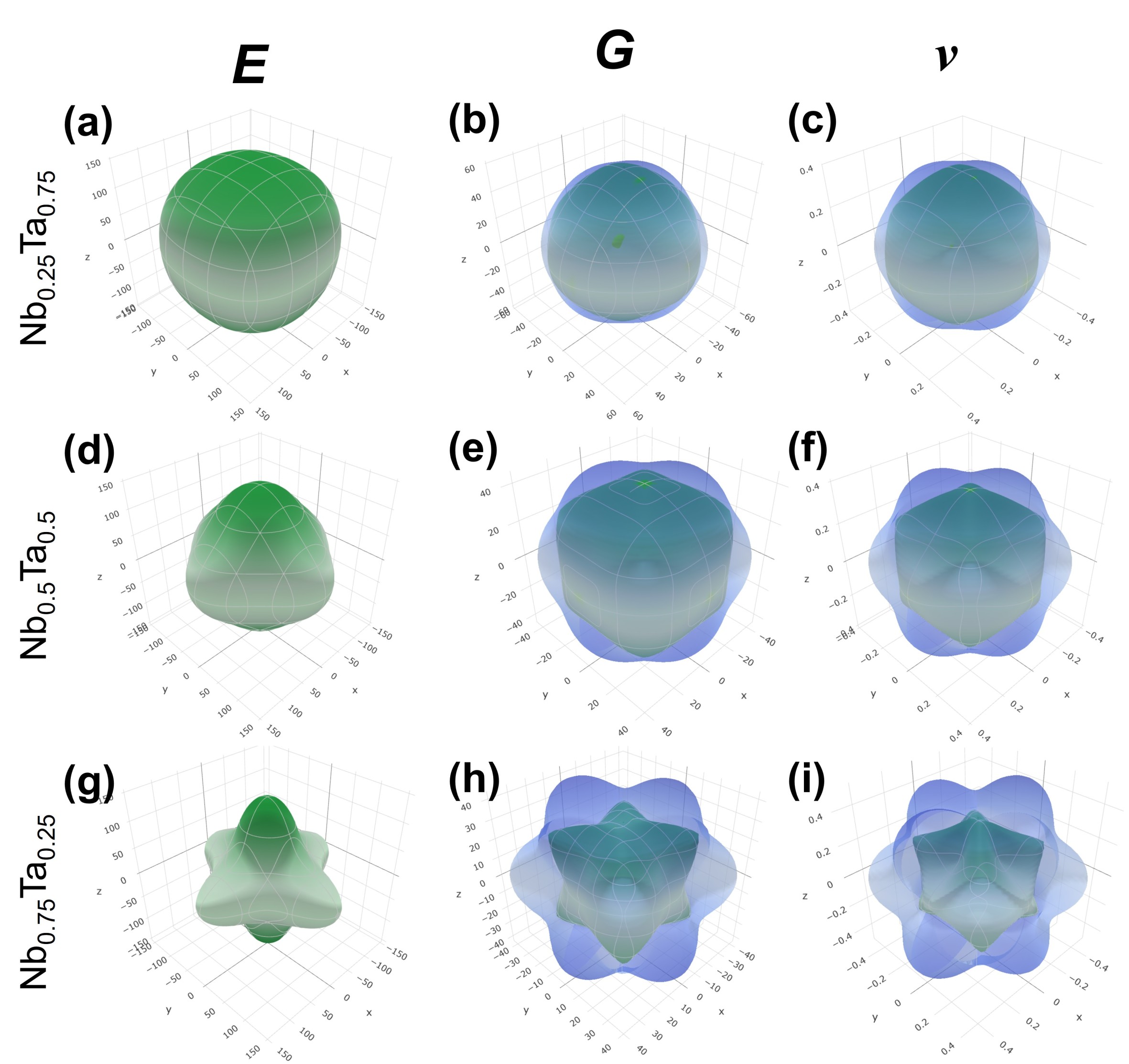}
\caption{\label{fig:anisotropy_NbTa} Anisotropy plots of $G$ (GPa), $E$ (GPa) and $\nu$ for: (a)-(c)  $\rm{Nb_{0.25}Ta_{0.75}}$;\, (d)-(f)  $\rm{Nb_{0.5}Ta_{0.5}}$;\, and (g)-(h) $\rm{Nb_{0.75}Ta_{0.25}}$.  Plots generated using the ELATE tool \cite{gaillac2016elate}. The maximum and minimum values of $G$, $E$ and $\nu$ are represented as transparent blue and solid green surfaces respectively.}
\end{figure*}

\begin{figure*}[htbp]
\centering
\includegraphics[width=1.0\linewidth]{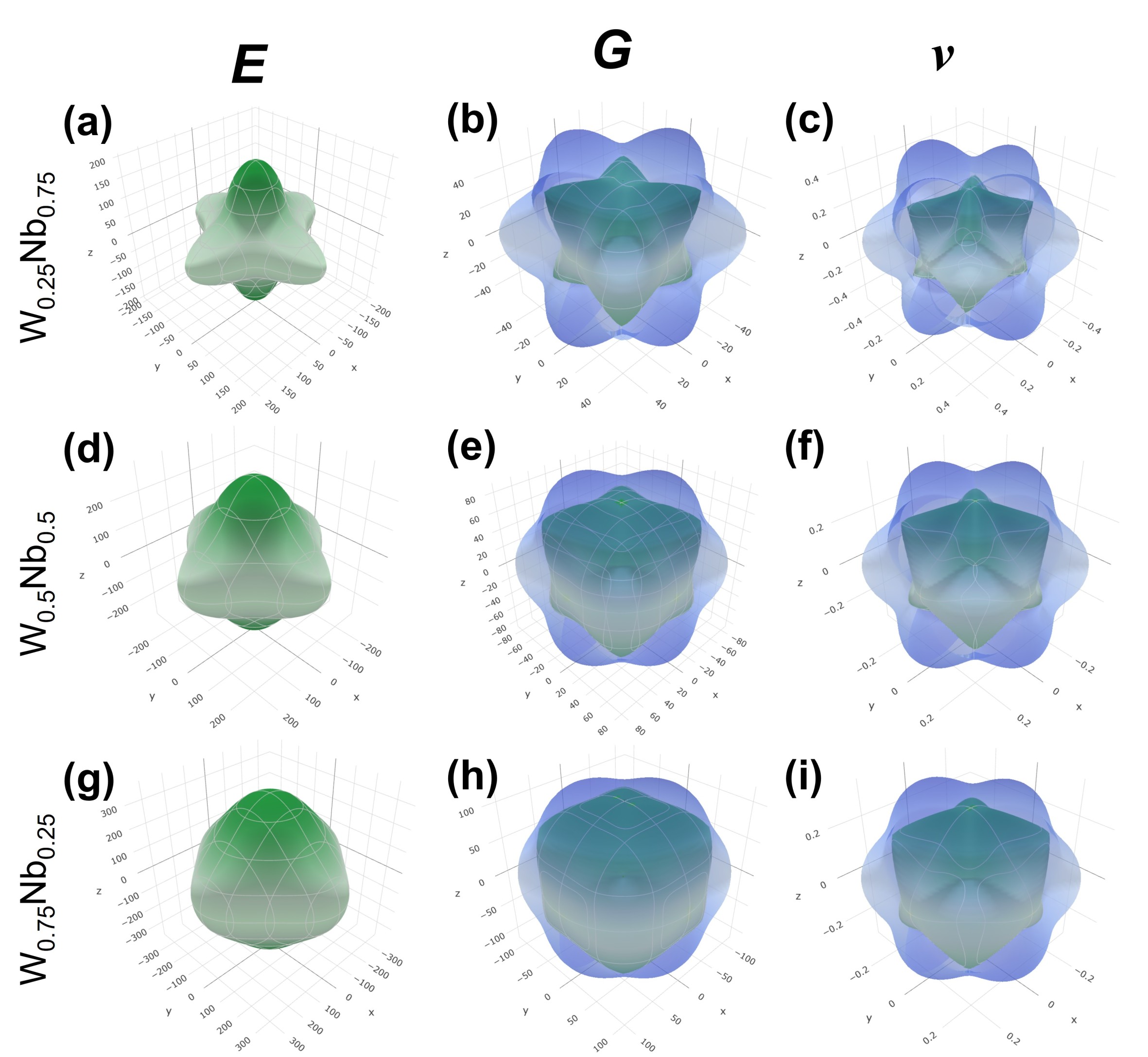}
\caption{\label{fig:anisotropy_WNb} Anisotropy plots of $G$ (GPa), $E$ (GPa) and $\nu$ for: (a)-(c)  $\rm{W_{0.25}Nb_{0.75}}$;\, (d)-(f)  $\rm{W_{0.5}Nb_{0.5}}$; and (g)-(h) $\rm{W_{0.75}Nb_{0.25}}$.  Plots generated using the ELATE tool \cite{gaillac2016elate}. The maximum and minimum values of $G$, $E$ and $\nu$ are represented as transparent blue and solid green surfaces respectively.}
\end{figure*}

\begin{figure*}[htbp]
\centering
\includegraphics[width=1.0\linewidth]{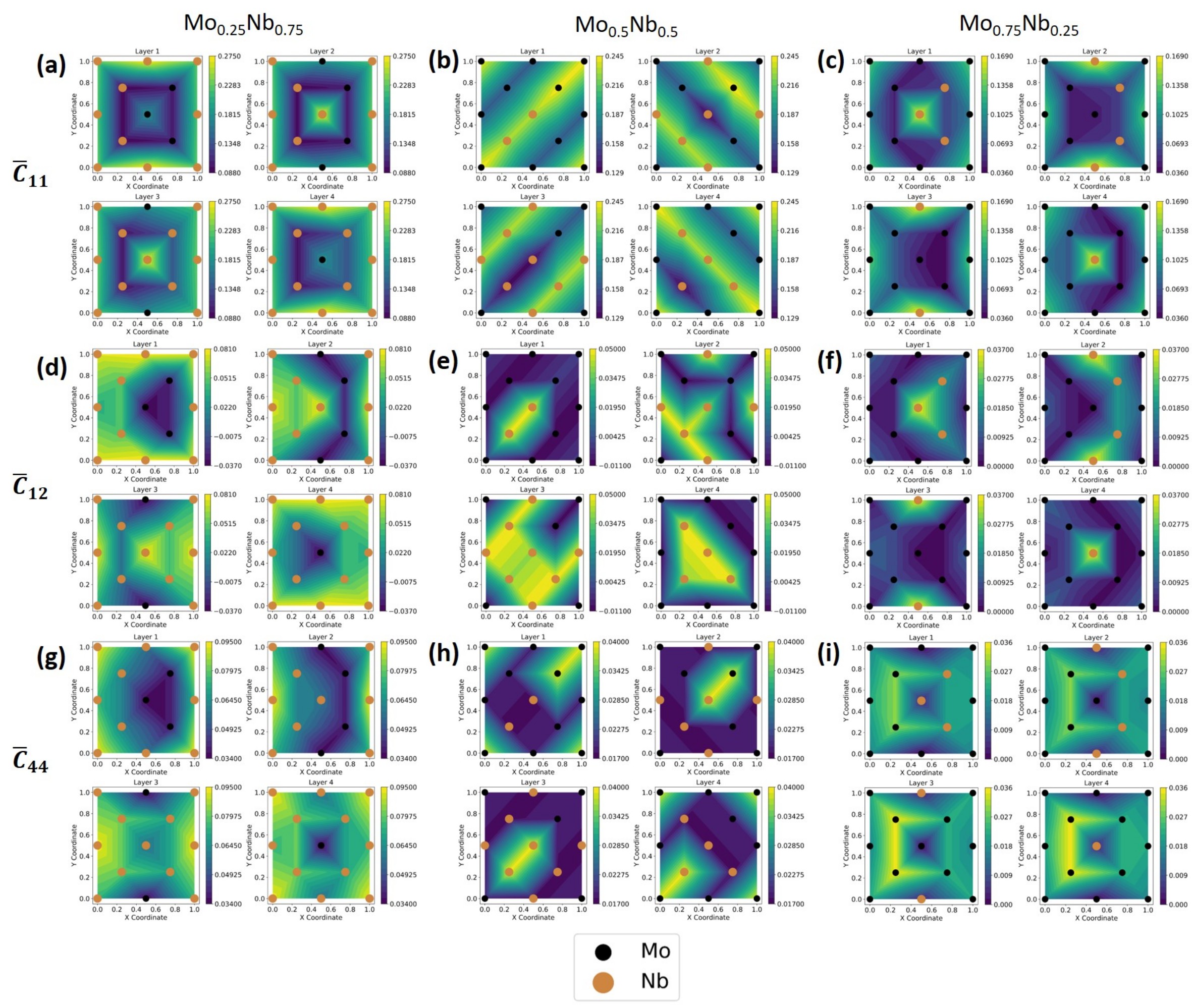}
\caption{\label{fig:forces_MoNb_1} Relaxation field (values in GPa) constructed as described in text for $\rm{Mo_{0.25}Nb_{0.75}}$, $\rm{Mo_{0.5}Nb_{0.5}}$, and $\rm{Mo_{0.75}Nb_{0.25}}$. (a)-(c): $\bar{C}_{11}$,\, (d)-(f): $\bar{C}_{12}$ and (g)-(i): $\bar{C}_{44}$. The color bar is set to range from minimum to maximum relaxation values for all atoms. $x$ and $y$ coordinates correspond to reduced coordinates in the 3D supercell.}
\end{figure*}

\begin{figure*}[htbp]
\centering
\includegraphics[width=1.0\linewidth]{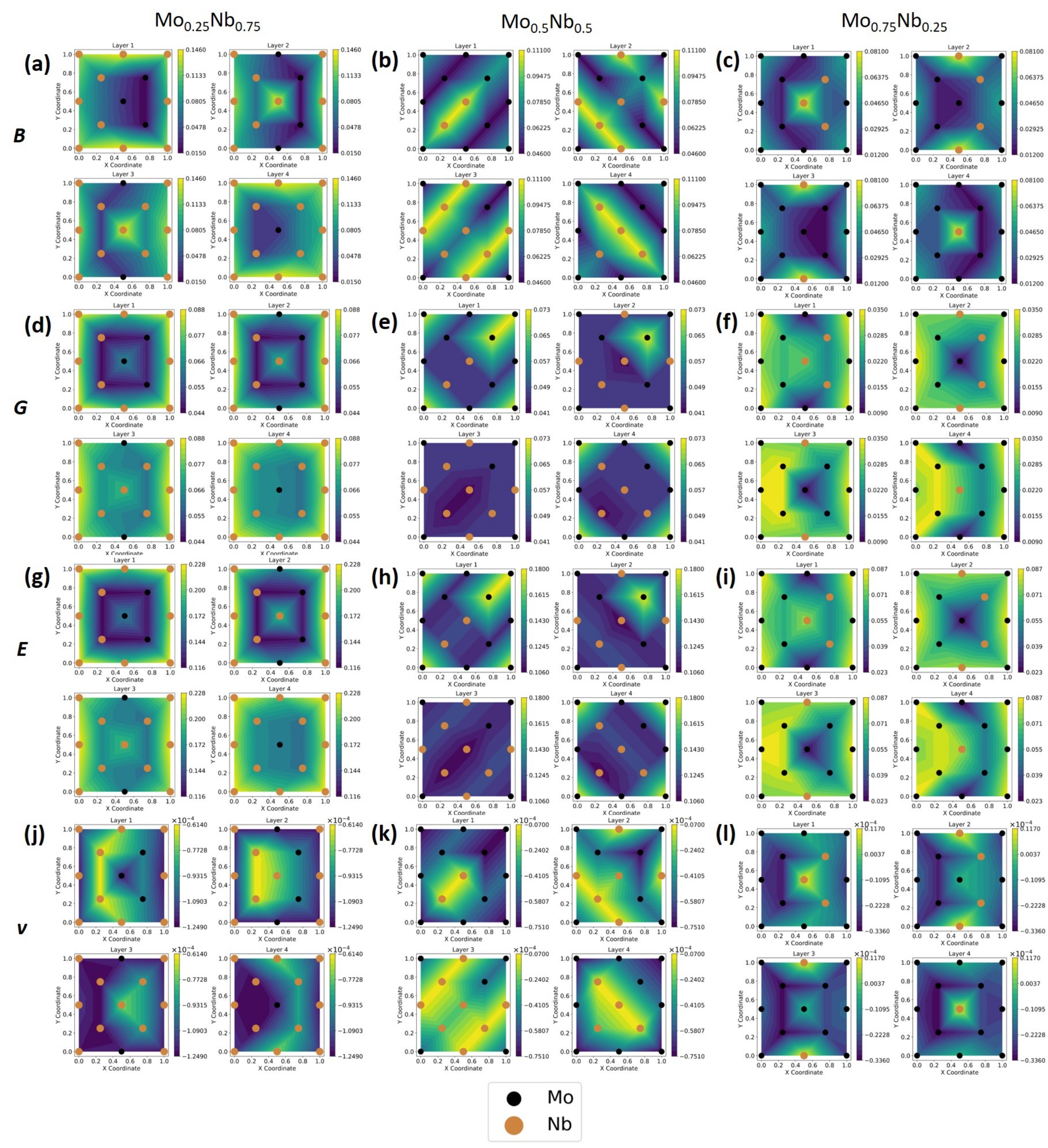}
\caption{\label{fig:forces_MoNb_2} Relaxation field (values in GPa) constructed as described in text for $\rm{Mo_{0.25}Nb_{0.75}}$, $\rm{Mo_{0.5}Nb_{0.5}}$, and $\rm{Mo_{0.75}Nb_{0.25}}$. (a)-(c): $B$,\, (d)-(f): $G$,\, (g)-(i): $E$ and (j)-(l): $\nu$. The color bar is set to range from minimum to maximum relaxation values for all atoms. $x$ and $y$ coordinates correspond to reduced coordinates in the 3D supercell.}
\end{figure*}

\begin{figure*}[htbp]
\centering
\includegraphics[width=1.0\linewidth]{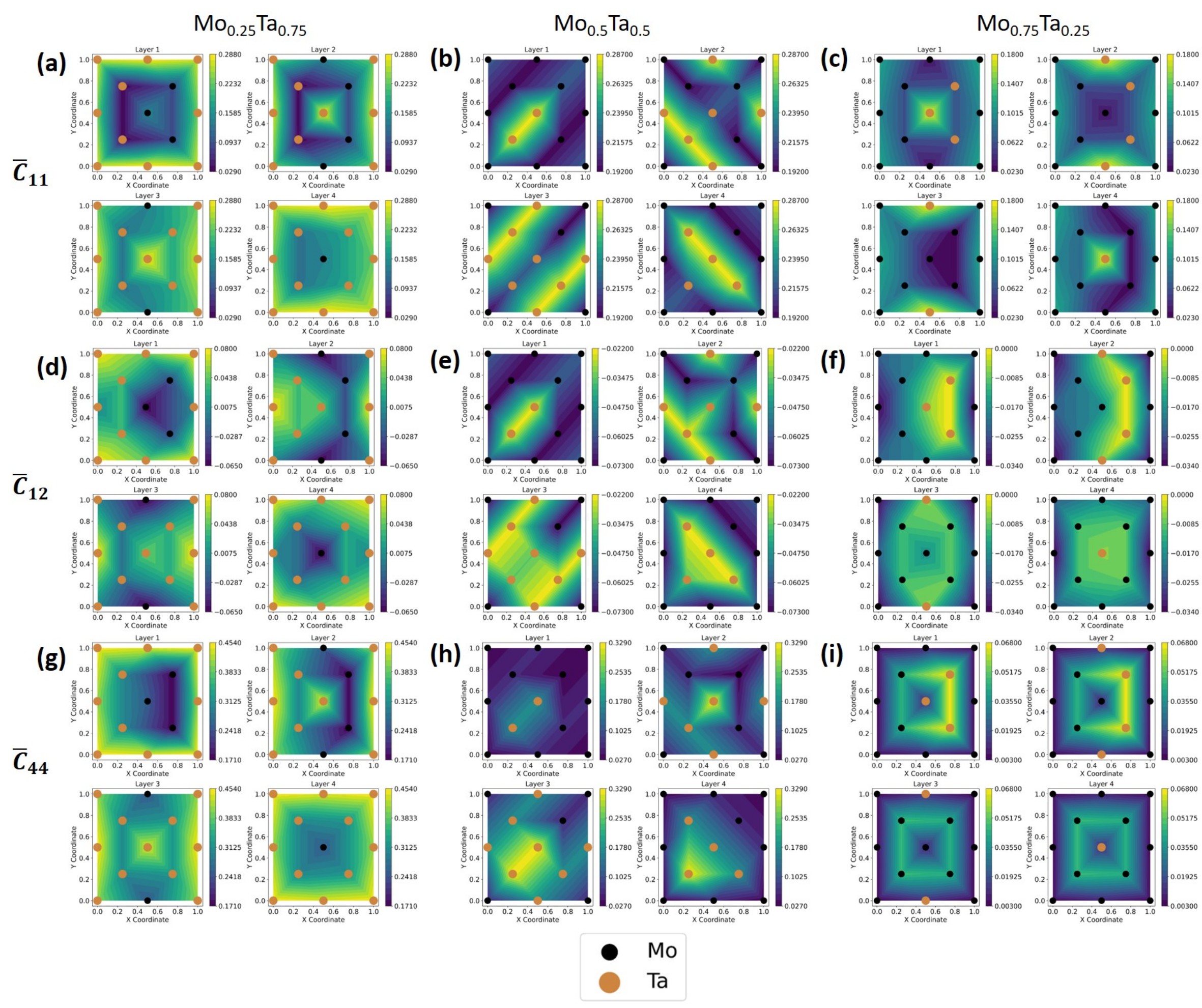}
\caption{\label{fig:forces_MoTa_1} Relaxation field (values in GPa) constructed as described in text for $\rm{Mo_{0.25}Ta_{0.75}}$, $\rm{Mo_{0.5}Ta_{0.5}}$, and $\rm{Mo_{0.75}Ta_{0.25}}$. (a)-(c): $\bar{C}_{11}$,\, (d)-(f): $\bar{C}_{12}$ and (g)-(i): $\bar{C}_{44}$. The color bar is set to range from minimum to maximum relaxation values for all atoms. $x$ and $y$ coordinates correspond to reduced coordinates in the 3D supercell.}
\end{figure*}

\begin{figure*}[htbp]
\centering
\includegraphics[width=1.0\linewidth]{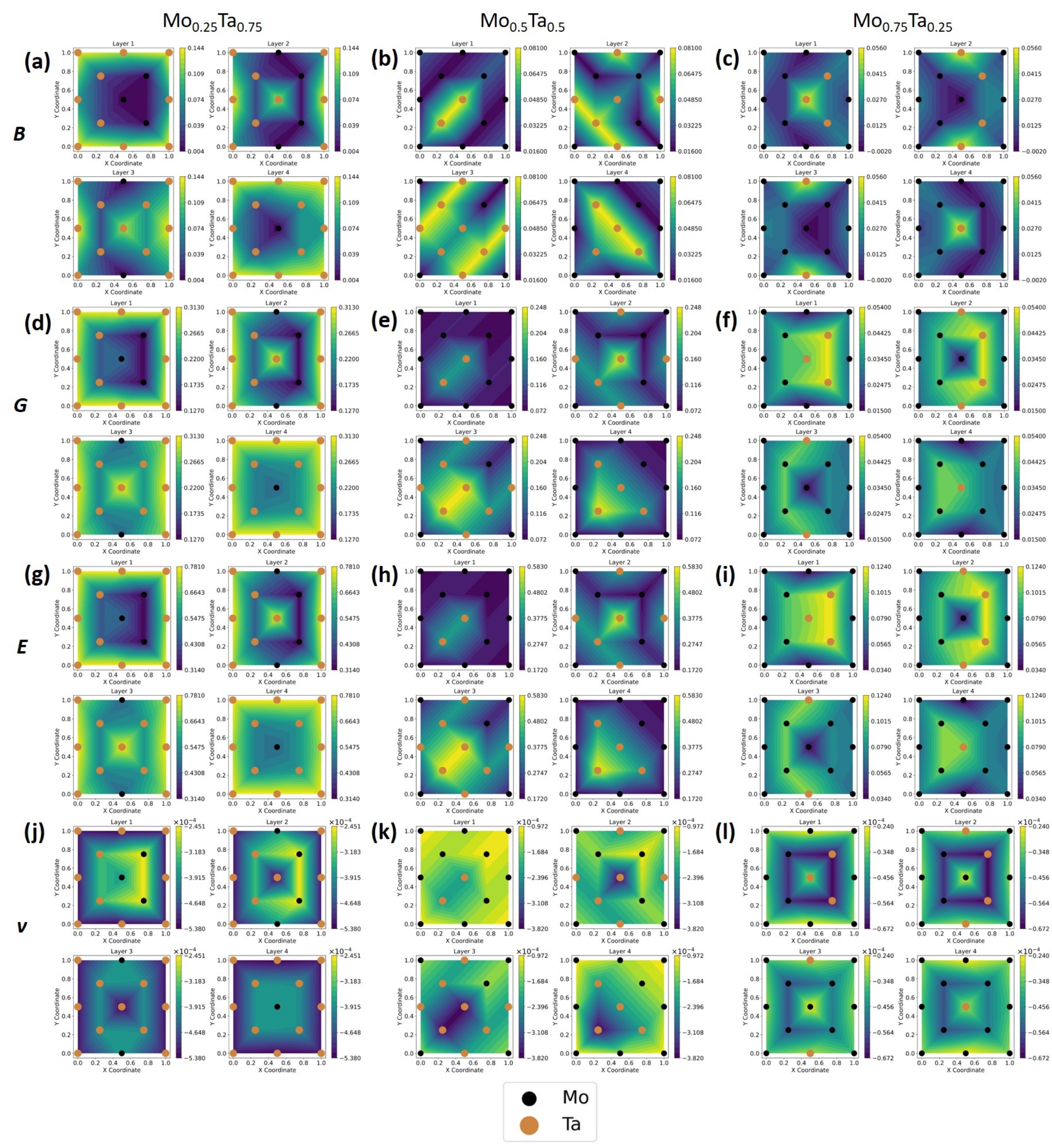}
\caption{\label{fig:forces_MoTa_2} Relaxation field (values in GPa) constructed as described in text for $\rm{Mo_{0.25}Ta_{0.75}}$, $\rm{Mo_{0.5}Ta_{0.5}}$, and $\rm{Mo_{0.75}Ta_{0.25}}$. (a)-(c): $B$,\, (d)-(f): $G$,\, (g)-(i): $E$ and (j)-(l): $\nu$. The color bar is set to range from minimum to maximum relaxation values for all atoms. $x$ and $y$ coordinates correspond to reduced coordinates in the 3D supercell.}
\end{figure*}

\begin{figure*}[htbp]
\centering
\includegraphics[width=1.0\linewidth]{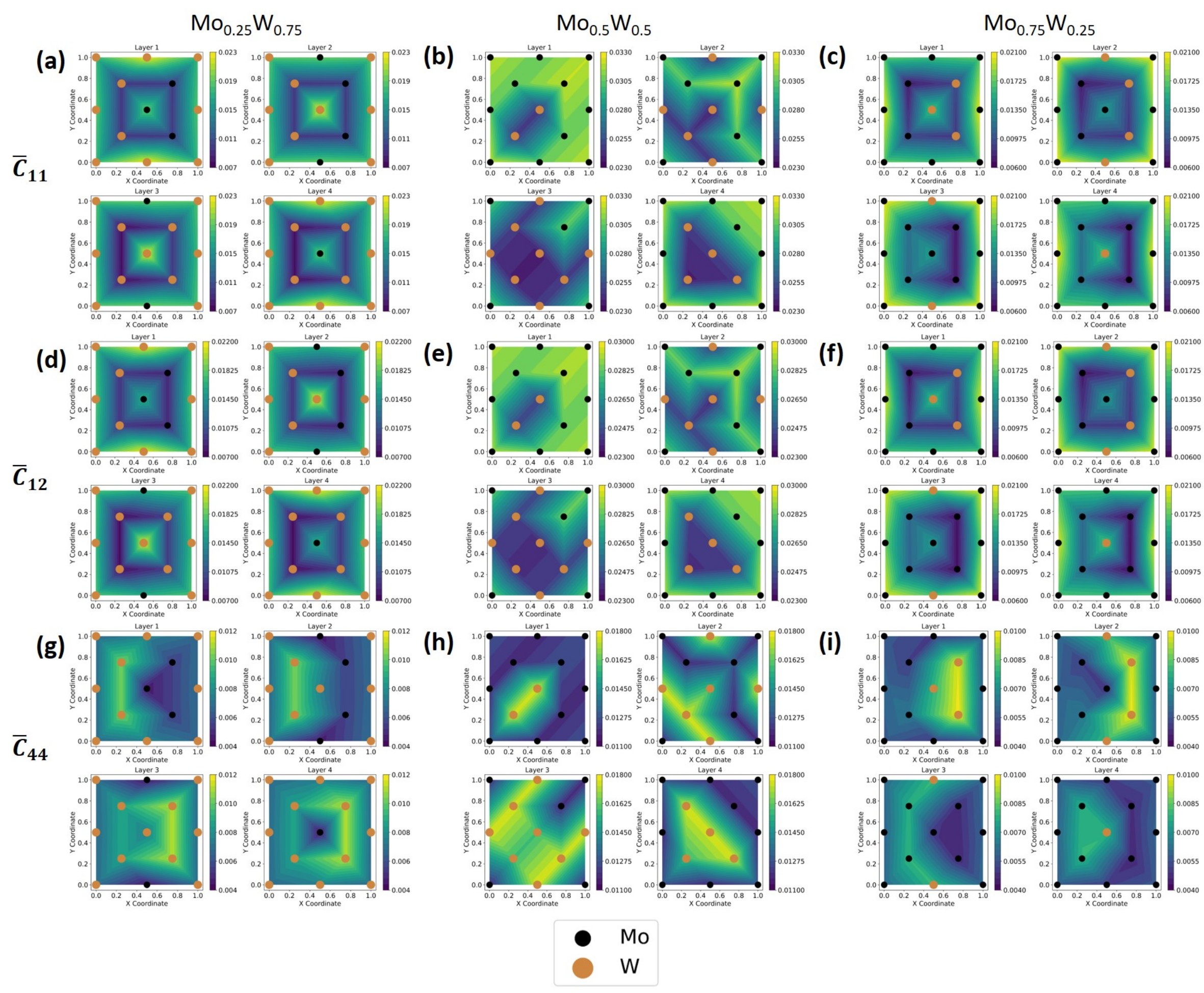}
\caption{\label{fig:forces_MoW_1} Relaxation field (values in GPa) constructed as described in text for $\rm{Mo_{0.25}W_{0.75}}$, $\rm{Mo_{0.5}W_{0.5}}$, and $\rm{Mo_{0.75}W_{0.25}}$. (a)-(c): $\bar{C}_{11}$,\, (d)-(f): $\bar{C}_{12}$ and (g)-(i): $\bar{C}_{44}$. The color bar is set to range from minimum to maximum relaxation values for all atoms. $x$ and $y$ coordinates correspond to reduced coordinates in the 3D supercell.}
\end{figure*}

\begin{figure*}[htbp]
\centering
\includegraphics[width=1.0\linewidth]{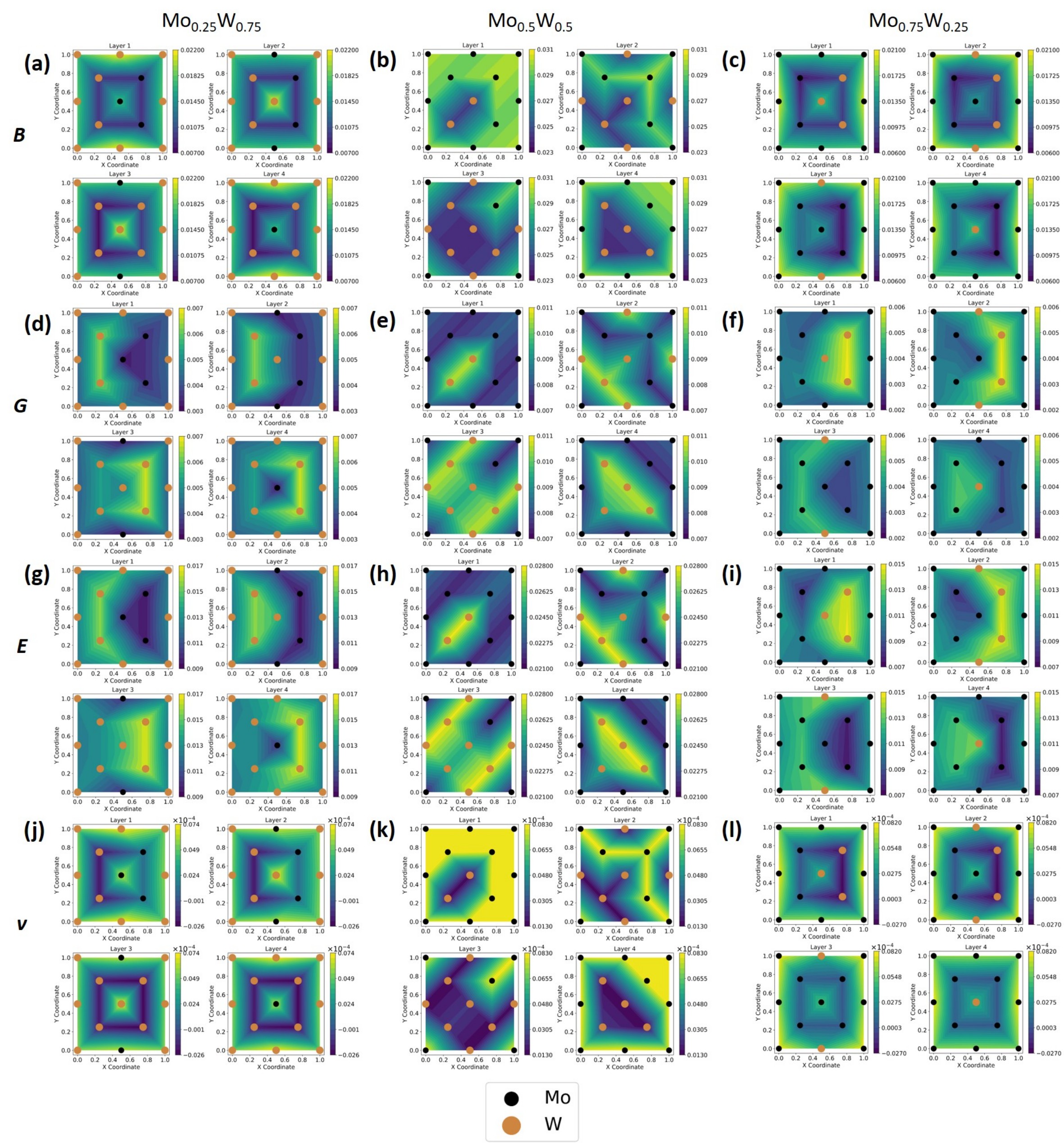}
\caption{\label{fig:forces_MoW_2} Relaxation field (values in GPa) constructed as described in text for $\rm{Mo_{0.25}W_{0.75}}$, $\rm{Mo_{0.5}W_{0.5}}$, and $\rm{Mo_{0.75}W_{0.25}}$. (a)-(c): $B$,\, (d)-(f): $G$,\, (g)-(i): $E$ and (j)-(l): $\nu$. The color bar is set to range from minimum to maximum relaxation values for all atoms. $x$ and $y$ coordinates correspond to reduced coordinates in the 3D supercell.}
\end{figure*}

\begin{figure*}[htbp]
\centering
\includegraphics[width=1.0\linewidth]{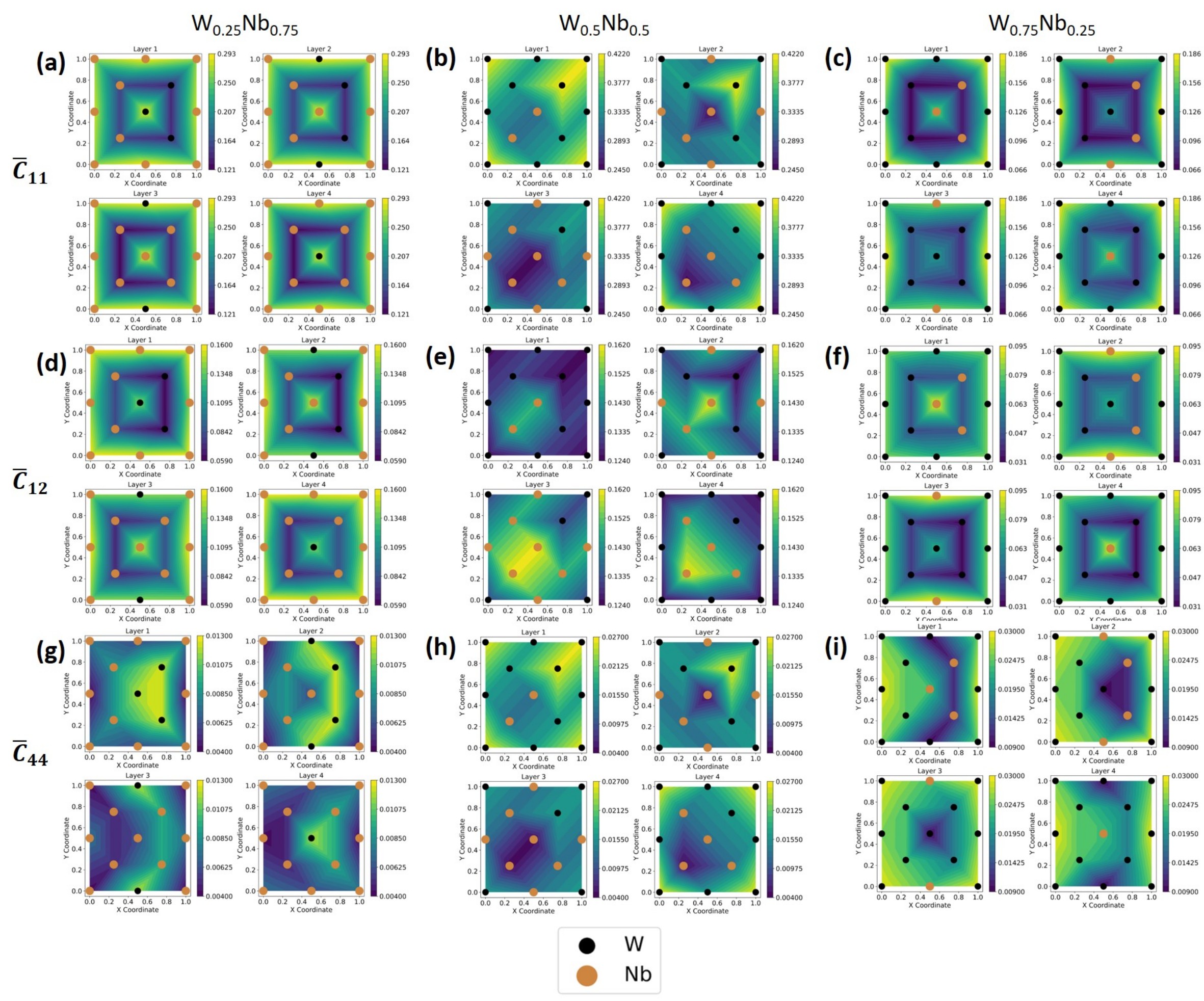}
\caption{\label{fig:forces_WNb_1} Relaxation field (values in GPa) constructed as described in text for $\rm{W_{0.25}Nb_{0.75}}$, $\rm{W_{0.5}Nb_{0.5}}$, and $\rm{W_{0.75}Nb_{0.25}}$. (a)-(c): $\bar{C}_{11}$,\, (d)-(f): $\bar{C}_{12}$ and (g)-(i): $\bar{C}_{44}$. The color bar is set to range from minimum to maximum relaxation values for all atoms. $x$ and $y$ coordinates correspond to reduced coordinates in the 3D supercell.}
\end{figure*}

\begin{figure*}[htbp]
\centering
\includegraphics[width=1.0\linewidth]{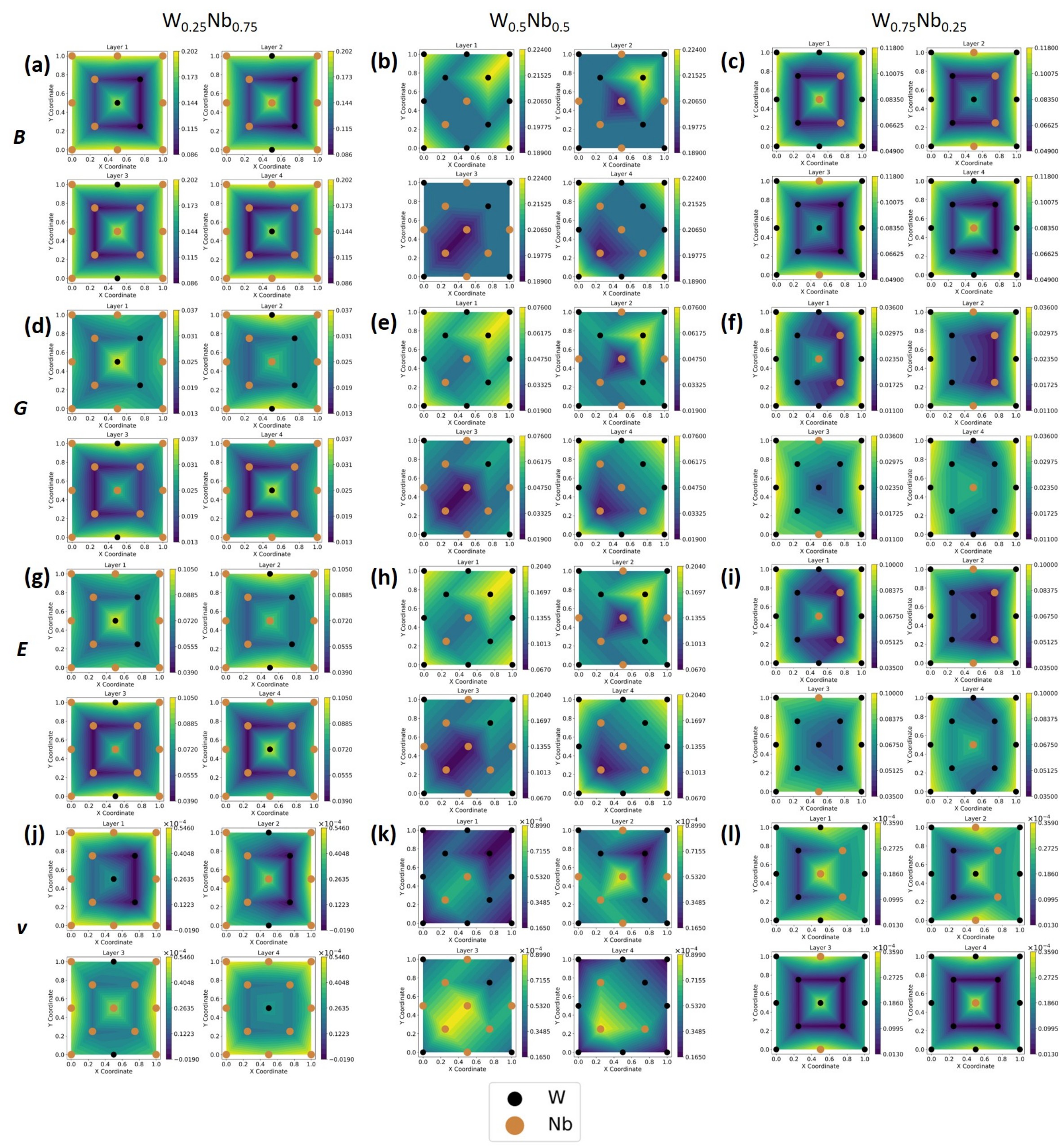}
\caption{\label{fig:forces_WNb_2} Relaxation field (values in GPa) constructed as described in text for $\rm{W_{0.25}Nb_{0.75}}$, $\rm{W_{0.5}Nb_{0.5}}$, and $\rm{W_{0.75}Nb_{0.25}}$. (a)-(c): $B$,\, (d)-(f): $G$,\, (g)-(i): $E$ and (j)-(l): $\nu$. The color bar is set to range from minimum to maximum relaxation values for all atoms. $x$ and $y$ coordinates correspond to reduced coordinates in the 3D supercell.}
\end{figure*}

The following figures provide additional computational results for binaries not discussed in the main text. Figures~\ref{fig:anisotropy_MoTa}--\ref{fig:anisotropy_WNb} illustrate the spatial dependence of $G$ (GPa), $E$ (GPa), and $\nu$ for MoTa, MoW, NbTa and WNb, as a function of composition. Figures~\ref{fig:forces_MoNb_1}--\ref{fig:forces_WNb_2} illustrate the nuclear-relaxation fields for $\bar{C}_{11}$, $\bar{C}_{12}$, $\bar{C}_{44}$, $B$, $G$, $E$ and $\nu$ as a function of composition for MoNb, MoTa, MoW and WNb.

\end{document}